\documentclass{siamart250211}

\usepackage{lipsum}
\usepackage{amsfonts}
\usepackage{graphicx}
\usepackage{epstopdf}
\usepackage{algorithmic}
\ifpdf
  \DeclareGraphicsExtensions{.eps,.pdf,.png,.jpg}
\else
  \DeclareGraphicsExtensions{.eps}
\fi
\newsiamremark{remark}{Remark}
\newsiamremark{hypothesis}{Hypothesis}
\crefname{hypothesis}{Hypothesis}{Hypotheses}
\newsiamthm{claim}{Claim}
\newsiamremark{fact}{Fact}
\crefname{fact}{Fact}{Facts}

\usepackage[normalem]{ulem}

\renewcommand{\(}{\left(}
\renewcommand{\)}{\right)}
\renewcommand{\[}{\left\lbrack}
\renewcommand{\]}{\right\rbrack}

\newcommand{\lbk}{\left\lbrace}
\newcommand{\rbk}{\right\rbrace}

\newcommand{\lnm}{\left\Vert}
\newcommand{\rnm}{\right\Vert}
\newcommand{\lmdl}{\left\vert}
\newcommand{\rmdl}{\right\vert}

\newcommand{\Id}{\textup{Id}} 
\newcommand{\tp}{^{T}} 
\newcommand{\kB}{k _{\textup{B}}} 
\DeclareMathOperator{\rank}{rank}

\newcommand{\bigO}[1]{\mathcal{O} \( {#1} \)} 

\newcommand{\mypower}[2]{%
  \ifthenelse{
    \( \equal{#2}{0} \)
  }{}{%
    \ifthenelse{
      \( \equal{#2}{1} \)
    }{#1}{{#1} ^{#2}}
  }
}
\newcommand{\mykgms}[3]{%
  \ifthenelse{
    \( \equal{#1}{0} \) \AND \( \equal{#2}{0} \) \AND \( \equal{#3}{0} \)
  }{%
    1}{}
  \ifthenelse{
    \( \equal{#1}{0} \) \AND \( \equal{#2}{0} \) \AND \( \NOT \equal{#3}{0} \)
  }{%
    \mypower{\textup{s}}{#3}}{}
  \ifthenelse{
    \( \equal{#1}{0} \) \AND \( \NOT \equal{#2}{0} \) \AND \( \equal{#3}{0} \)
  }{%
    \mypower{\textup{m}}{#2}}{}
  \ifthenelse{
    \( \equal{#1}{0} \) \AND \( \NOT \equal{#2}{0} \) \AND \( \NOT \equal{#3}{0} \)
  }{%
    \mypower{\textup{m}}{#2} \cdot \mypower{\textup{s}}{#3}}{}
  \ifthenelse{
    \( \NOT \equal{#1}{0} \) \AND \( \equal{#2}{0} \) \AND \( \equal{#3}{0} \)
  }{%
    \mypower{\textup{kg}}{#1}}{}
  \ifthenelse{
    \( \NOT \equal{#1}{0} \) \AND \( \equal{#2}{0} \) \AND \( \NOT \equal{#3}{0} \)
  }{%
    \mypower{\textup{kg}}{#1} \cdot \mypower{\textup{s}}{#3}}{}
  \ifthenelse{
    \( \NOT \equal{#1}{0} \) \AND \( \NOT \equal{#2}{0} \) \AND \( \equal{#3}{0} \)
  }{%
    \mypower{\textup{kg}}{#1} \cdot \mypower{\textup{m}}{#2}}{}
  \ifthenelse{
    \( \NOT \equal{#1}{0} \) \AND \( \NOT \equal{#2}{0} \) \AND \( \NOT \equal{#3}{0} \)
  }{%
    \mypower{\textup{kg}}{#1} \cdot \mypower{\textup{m}}{#2} \cdot \mypower{\textup{s}}{#3}}{}
}
\newcommand{\e}[1]{\times 10 ^{#1}}

\headers{Reduced-Order Deterministic-Particle-Based Scheme}{Lidong Fang, Xuelian Bao, Zilong Song, Shixin Xu, Huaxiong Huang}

\title{Reduced-Order Variational Deterministic-Particle-Based Scheme for Fokker-Planck Equations in Microscopic Polymer Dynamics
  \thanks{Submitted to the editors in August 2025. 
  }
}

\author{Lidong Fang
  \thanks{School of Mathematics, Shanghai University of Finance and Economics, Shanghai, 200433, China (\email{fanglidong@sufe.edu.cn}).}
\and Xuelian Bao
  \thanks{School of Mathematics, South China University of Technology, Guangzhou, 510641, China (\email{baoxuelian@scut.edu.cn}).}
\and Zilong Song
  \thanks{Department of Mathematics and Statistics, Utah State University, Logan, 84322, USA (\email{zilong.song@usu.edu}).}
\and Shixin Xu
  \thanks{Zu Chongzhi Center, Duke Kunshan University, Suzhou, 215316, China (\email{shixin.xu@dukekunshan.edu.cn}).}
\and Huaxiong Huang
  \thanks{Zu Chongzhi Center, Duke Kunshan University, Suzhou, 215316, China; Department of Mathematics and Statistics, York University, Toronto, M3J 1P3, Canada (\email{huaxiong.huang@dukekunshan.edu.cn}). Corresponding author.}
}

\ifpdf
\hypersetup{
  pdftitle={VDS POD-MOR},
  pdfauthor={Fang, et al}
}
\fi

\begin{document}

\maketitle

\begin{abstract}
  This study proposes an acceleration technique for the computational challenges in extending the variational deterministic-particle-based scheme (VDS) [Bao et al., Journal of Computational Physics 522 (2025) 113589] to 3D complex fluid simulations with multi-bead polymers. 
  While the original VDS effectively captures configuration space dynamics for 2D dumbbell polymers, its direct extensions reveal critical scalability limitations. The growing configuration space dimensionality necessitates prohibitively large particle ensembles to maintain distributional accuracy, so its quadratic computational cost scaling impedes practical applications. 
  In this paper, we develop a model reduction framework integrating proper orthogonal decomposition (POD) to speed up the computation of the VDS for microscopic Fokker-Planck equations.
  Numerical validation using bead-spring chain models in simple shear flow demonstrates that the computational efficiency of the reduced model increases systematically with molecular complexity. 
  The reduced-order model introduces about $6\%$ relative error in predicting the dynamics while requiring only about $6\%$ of the original computational time for $4$-bead chain polymers, where the relative numerical error of the reference dynamics is about $5\% \sim 10\%$, and the degrees of freedom can be reduced significantly to about $0.1\%$ of the original model, which means the low-dimensional structure is found by POD.
  This establishes a practical pathway for multiscale and complex fluid simulations.
\end{abstract}



\section{Introduction}
\label{sec:introduction}

\subsection{Background}

Micro-macro modeling of dilute polymeric fluids bridges macroscopic hydrodynamics with microscopic polymer dynamics through coupled Navier-Stokes and Fokker-Planck equations \cite{bird1987dynamics, laso1993calculation, le2012micro, fang2022deepn}. 


However, direct computation remains prohibitive for practical applications, primarily due to the high-dimensional configuration space of polymer molecules. While stochastic methods (e.g., CONNFFESSIT \cite{ottinger2012stochastic}) alleviate this via Brownian dynamics, their inherent noise necessitates extensive ensemble averaging, which still requires large computation.

\subsection{The VDS Scheme and its Limitation}

The variational deterministic-particle-based scheme (VDS) is inspired by the discrete energetic variational approach \cite{wang2021two}, and is developed in \cite{bao2024micro, bao2025deterministic} to solve the microscopic Fokker-Planck equation of dilute polymeric fluids in micro-macro coupling models.

The VDS replaces stochastic trajectories with a regularized empirical measure for the conformation distribution $f \( \mathbf{x}, \mathbf{q}, t \)$ through kernel smoothing. 
The particle dynamics is derived from an energy-dissipation law, preserving thermodynamic consistency.

Coupling with macroscopic Navier-Stokes equations, VDS is validated numerically by the 2D Hookean dumbbell model and the 2D FENE dumbbell model under different flows \cite{bao2025deterministic} and the 2D dumbbell model with irreversible bond-breaking potential energy and modified Morse potentials under simple shear flow \cite{bao2024micro}.

Despite its advantages --- notably eliminating statistical noise --- VDS faces scalability challenges when extended to 3D multi-bead polymers for more practical applications, the dimension of the polymer configuration space is higher. As a consequence, estimating the polymer number distribution through discrete representative particles requires more particles to maintain the approximation accuracy.
This results in an issue of computational cost since pairwise kernel evaluations yield $\bigO{P ^{2}}$ operations per time step when the molecular complexity amplifies particle counts $P$ to maintain accuracy \cite{bao2025micro}. 
That is the reason why most of deterministic schemes focus on 2D dumbbell models, while most of the recent works on multi-bead models are based on the stochastic methods \cite{bao2025deterministic}.

On the other hand, as in many applications, the configuration space often lies in lower dimensional spaces, which motivates the search for more efficient computational method using model-order reduction (MOR).
Near thermal equilibrium, we are allowed to use dumbbells to approximate longer chains. Since most interesting phenomena, such as viscoelasticity, happen away from thermal equilibrium in complex fluids, a direct reduction of the number of beads does not work.
We need to use a more general MOR approach.

\subsection{POD-MOR}

MOR is important in engineering and science because it allows for efficient simulations by significantly lowering the degrees of freedom (DoFs) while keeping the reduction error small \cite{lassila2014model, lucia2004reduced, peherstorfer2015dynamic}. Physical simulations can be expensive due to their complexity, especially in cases like parameter studies or design optimization that require multiple simulations, where MOR is particularly helpful \cite{copeland2022reduced}.

The most common MOR method uses projection techniques \cite{benner2015survey}, which approximate state variables in a lower-dimensional space \cite{mojgani2017lagrangian}. A typical projection-based MOR has two parts: an offline phase that reduces dimensions by identifying the key modes in the dynamics and an online phase that calculates approximate solutions \cite{magargal2022lagrangian}. One leading method for offline dimensional reduction is Proper Orthogonal Decomposition (POD) method \cite{sirovich1987turbulence1, sirovich1987turbulence2, sirovich1987turbulence3, holmes1996turbulence} is best known in fluid computations \cite{lall2003structure, taira2017modal}, and POD is also known as the Karhunen–Lo\`eve expansion decomposition (KLD) \cite{loeve1945calcul, karhunen1946spektraltheorie} in signal processing community or the principle component analysis (PCA) \cite{pearson1901liii, hotelling1933analysis} in statistics community. Other offline methods like balanced truncation methods also have strong relation to the POD method \cite{lall2003structure, rowley2005model}. In the online phase, either Galerkin projection or Petrov-Galerkin projection can be used \cite{rowley2004model, rapun2010reduced, carlberg2017galerkin, choi2019space, carlberg2011efficient}. We use the term POD-MOR to refer to the POD-based MOR method with Galerkin projection.

As a data representation technique, POD finds the best lower-dimensional approximating subspace (orthogonal basis) to a given set of empirical data. More precisely, POD generates subspace basis $\mathbf{U} \in \mathbb{R} ^{F \times R}$ from snapshots $\mathbf{X} = \( \mathbf{w} ^{1}, \cdots, \mathbf{w} ^{D} \) \in \mathbb{R} ^{F \times D}$. Here $F$ and $R$ are the DoFs of the full and reduced system, and $D$ is the size of the dataset.
In fact, POD solves the best low-rank approximation
\begin{align*}
  \min _{\rank \( \tilde{\mathbf{X}} \) = R} \lnm \mathbf{X} - \tilde{\mathbf{X}} \rnm _{\textup{Frobenius}}.
\end{align*}
Equivalently, POD solves the singular value decomposition (SVD) of the data $\mathbf{X}$,
\begin{align*}
  \mathbf{X} = \hat{\mathbf{U}} \cdot \hat{\boldsymbol{\Sigma}} \cdot \hat{\mathbf{V}} \tp,
  \quad
  \tilde{\mathbf{X}} = \hat{\mathbf{U}} _{:, 1 : R} \cdot \hat{\boldsymbol{\Sigma}} _{1 : R, 1 : R} \cdot \[ \hat{\mathbf{V}} _{:, 1 : R} \] \tp,
  \quad
  \mathbf{U} := \hat{\mathbf{U}} _{:, 1 : R},
\end{align*}
or solves the eigenvalue decomposition of the covariance matrix $\mathbf{X} \cdot \mathbf{X} \tp$,
\begin{align*}
  \mathbf{X} \cdot \mathbf{X} \tp = \hat{\mathbf{U}} \cdot \hat{\boldsymbol{\Sigma}} \cdot \hat{\mathbf{U}} \tp,
  \quad
  \mathbf{U} := \hat{\mathbf{U}} _{:, 1 : R},
\end{align*}
where the notation $1 : R$ indicates the set of indices $\lbk 1, 2, \cdots, R \rbk$.

The POD method can automatically find the most important spatial modes in a dataset, and was originally used as a data representation technique, and later, POD is used for model reduction of dynamical systems with the system trajectories obtained via experiments, numerical simulations, or analytical derivations \cite{rathinam2003new}. By capturing the key features of the system, POD allows for the representation of complex phenomena with far fewer DoFs. This not only reduces computational costs but also improves understanding of the underlying physics \cite{chatterjee2000introduction, berkooz1993proper, kerschen2005method}. The effectiveness of POD is especially clear when applied to the heat equation on a fixed grid. This method works well in this context due to the fast decay of high-frequency modes, so that the number of DoFs (i.e., the number of important low-frequency modes) can be effectively reduced. Another example is the complex friction stir welding problem where POD can be applied to achieve MOR and improve computational efficiency \cite{cao2022machine, fang0000proper1}.

\subsection{Contribution}


In this paper, we study an extension of the original VDS from 2D systems to 3D systems and from dumbbell polymers to multi-bead polymers, which is much closer to real-world applications.
Due to the inhibitive computational cost, it is impractical to solve VDS in its original form, so we develop a POD-MOR approach to speed up the computation of the VDS for microscopic Fokker-Planck equations. 
Numerical validation using bead-spring chain models in simple shear flow demonstrates that the computational efficiency of the reduced model increases systematically with molecular complexity.
The reduced-order model introduces about $6\%$ relative error in predicting the dynamics while requiring only about $6\%$ of the original computational time for $4$-bead chain polymers, where the error benchmark, i.e., the relative numerical error of the reference dynamics is about $5\% \sim 10\%$. At the same time, the degrees of freedom can be reduced significantly to about $0.1\%$ of the original model, which means the low-dimensional structure is found by POD.

While previous work \cite{fang0000proper2} proved the concept on a toy model --- a 1D bead-spring model --- our study further validates the POD's feasibility on Lagrangian systems and demonstrates the scalability of this approach in real-world and larger-scale settings.

This method is able to extend to solve the coupled system of Fokker-Planck equations and the Navier-Stokes equations for modeling the full hydrodynamics of the polymeric fluid, which means this work establishes a practical pathway for multiscale and complex fluid simulations. We will explore it in the future work.

\subsection{Outline}

The paper is organized as follows: \cref{sec:model} details the extension of the original VDS and the POD-MOR approach, \cref{sec:experiments} presents numerical benchmarks, and \cref{sec:conclusions} concludes this paper.

\section{Model}
\label{sec:model}

\subsection{Bead-Bond Model and Fokker-Planck Equation}

In modeling dilute polymeric fluids with the bead-bond chain model at the microscopic level, one polymer molecule is represented by a linear chain of $N$ beads connected by $N - 1$ springs \cite{lin2007micro, ottinger2012stochastic}.
The molecular configuration is characterized by end-to-end bond vectors $\mathbf{q} = \( \mathbf{q} _{1}, \cdots, \mathbf{q} _{N - 1} \) \in \mathbb{R} ^{\[ N - 1 \] \times d} \[ \mykgms{0}{1}{0} \]$, where $\mathbf{q} _{j} = \( q _{j, 1}, \cdots, q _{j, d} \) \in \mathbb{R} ^{d} \[ \mykgms{0}{1}{0} \]$, $j \in \lbk 1, \cdots, N - 1 \rbk$, and the spatial dimension is $d = 3$.
We denote $\mathbf{r} _{j} \in \mathbb{R} ^{d} \[ \mykgms{0}{1}{0} \]$ as the position vector of the $j$-th bead, $j \in \lbk 1, \cdots, N \rbk$, so that the bond vector $\mathbf{q} _{j} = \mathbf{r} _{j + 1} - \mathbf{r} _{j}$, $j \in \lbk 1, \cdots, N - 1 \rbk$.

The microscopic Fokker-Planck equation that describes the dynamics of the polymer configuration density function $f$ reads \cite{bird1987dynamics, laso1993calculation, le2012micro, fang2022deepn}, 
\begin{align}
  &
  \quad
  \partial _{t} f
  +
  \mathbf{u} \cdot \nabla _{\mathbf{x}} f
  \nonumber 
  \\
  & =
  - \sum _{j = 1} ^{N - 1} \nabla _{\mathbf{q} _{j}}
  \cdot
  \[ 
  \[ \nabla _{\mathbf{x}} \mathbf{u} \] \tp \cdot \mathbf{q} _{j} f
  -
  \frac{1}{\zeta} \sum _{k = 1} ^{N - 1} A _{jk} \nabla _{\mathbf{q} _{k}} \Psi \( \mathbf{q} \) f
  -
  \frac{\kB T}{\zeta} \sum _{k = 1} ^{N - 1} A _{jk} \nabla _{\mathbf{q} _{k}} f
  \]
  .
  \label{eq:model_FP}
\end{align}
Here the polymer configuration density function $f = f \( \mathbf{q}; \mathbf{x}, t \) \in \mathbb{R} \[ \mykgms{0}{-\[ N - 1 \] \times d}{0} \]$ represents the probability density in the configuration space at given space point $\mathbf{x} \in \mathbb{R} ^{d} \[ \mykgms{0}{1}{0} \]$ and given time $t \in \mathbb{R} \[ \mykgms{0}{0}{1} \]$, so that $\int f \( \mathbf{q}; \mathbf{x}, t \) \mathrm{d} \mathbf{q} \equiv 1$;
$\mathbf{u} = \mathbf{u} \( \mathbf{x}, t \) \in \mathbb{R} ^{d} \[ \mykgms{0}{1}{-1} \]$ is the macroscopic velocity field;
$\zeta \in \mathbb{R} \[ \mykgms{1}{0}{-1} \]$ is a friction constant related to the polymer relaxation time;
$\kB \in \mathbb{R} \[ \mykgms{1}{2}{-2} \cdot \textup{K} ^{-1} \]$ is the Boltzmann constant;
$T \in \mathbb{R} \[ \textup{K} \]$ is the absolute temperature;
$\Psi \( \mathbf{q} \) \in \mathbb{R} \[ \mykgms{1}{2}{-2} \]$ is the microscopic potential energy between beads that depends only on the current molecular configuration $\mathbf{q}$;
$\( A _{jk} \) _{j, k \in \lbk 1, \cdots, N - 1 \rbk} \in \mathbb{R}$ are the elements of the (dimensionless) Rouse matrix.

In this work, we focus on the Fokker-Planck equation (without coupling with the Navier-Stokes equation) with given velocity field $\mathbf{u} \( \mathbf{x}, t \)$ and the velocity gradient $\nabla _{\mathbf{x}} \mathbf{u}$. The case that fully coupled with the Navier-Stokes equation will be studied in the future work.

For simplicity, we choose the microscopic potential energy as the classical Hookean potential, that is,
\begin{align}
  \Psi \( \mathbf{q} \) = \sum _{j = 1} ^{N - 1} \Psi _{j} \( \mathbf{q} _{j} \) = \sum _{j = 1} ^{N - 1} \frac{H _{j}}{2} \lmdl \mathbf{q} _{j} \rmdl ^{2},
  \label{eq:model_Hookean}
\end{align}
where $\Psi _{j}$ is the potential energy on the $j$-th bond, and $H _{j} \in \mathbb{R} _{+} \[ \mykgms{1}{0}{-2} \]$ is the elastic constant on the $j$-th bond.

The polymer stress $\boldsymbol{\tau} = \boldsymbol{\tau} \( \mathbf{x}, t \) \in \mathbb{R} ^{d \times d} \[ \mykgms{1}{2-d}{-2} \]$ is defined as
\begin{align*}
  \boldsymbol{\tau} \( \mathbf{x}, t \)
  =
  n \( \mathbf{x}, t \)
  \int \sum _{j = 1} ^{N - 1} \mathbf{q} _{j} \nabla \Psi _{j} \( \mathbf{q} _{j} \) f \( \mathbf{q}; \mathbf{x}, t \) \mathrm{d} \mathbf{q},
\end{align*}
where $n = n \( \mathbf{x}, t \) \in \mathbb{R} \[ \mykgms{0}{-d}{0} \]$ is the polymer number density (number per unit volume).

Since we focus on chain polymer molecules, the elements of the Rouse matrix are defined as 
\begin{align*}
  A _{jk} = 2 \mathbf{1} _{j = k} - \mathbf{1} _{\lmdl j - k \rmdl = 1},
\end{align*}
where $\mathbf{1}$ is the characteristic function.

We notice that directly solving the microscopic Fokker-Planck equation \autoref{eq:model_FP} is computationally expansive when $N$ is not small.

\subsection{Variational Deterministic-Particle-Based Scheme}

One of the approaches to solve the microscopic Fokker-Planck equation \autoref{eq:model_FP} is VDS \cite{bao2025deterministic}, where the polymer configuration density function is represented by the distribution of a set of representative particles, and the evolution of the polymer configuration density function is transformed into a deterministic evolution of those particles.

Following the VDS, the deterministic particle approximation of the polymer configuration density function $f \( \mathbf{q}; \mathbf{x}, t \)$ at a given space point $\mathbf{x}$ and time $t$ requires a set of $P$ representative particles in the space of molecular configuration at position/configuration $\bar{\mathbf{q}} \( \mathbf{x}, t \) \in \mathbb{R} ^{\[ N - 1 \] \times P \times d} \[ \mykgms{0}{1}{0} \]$, and a smooth kernel function $K _{h}: \mathbb{R} ^{\[ N - 1 \] \times d} \[ \mykgms{0}{1}{0} \] \to \mathbb{R} \[ \mykgms{0}{- \[ N - 1 \] \times d}{0} \]$.

Here, the configuration of representative particles is $\bar{\mathbf{q}} = \( \bar{\mathbf{q}} _{k \( I \alpha \)} \)$, whereas the configuration of particle $I \in \lbk 1, \cdots, P \rbk$, bond $k \in \lbk 1, \cdots, N - 1 \rbk$, and spatial dimension $\alpha \in \lbk 1, \cdots, d \rbk$.


Furthermore, we denote $\mathbf{q} _{k :} \in \mathbb{R} ^{P d} \[ \mykgms{0}{1}{0} \]$ as the configuration vector of bond $k$ for all particles and all spatial dimensions, $\mathbf{q} _{I} := \mathbf{q} _{: \( I : \)} \in \mathbb{R} ^{\[ N - 1 \] d} \[ \mykgms{0}{1}{0} \]$ as the configuration vector of particle $I$ in all bonds and all spatial dimensions, $\mathbf{q} _{k, I} := \mathbf{q} _{k \( I : \)} \in \mathbb{R} ^{d} \[ \mykgms{0}{1}{0} \]$ as the configuration vector of $k$-th bond of particle $I$ in all spatial dimensions.

The deterministic particle approximation reads
\begin{align}
  f \( \mathbf{q}; \mathbf{x}, t \) \approx f _{P} \( \mathbf{q}; \mathbf{x}, t \) = \frac{1}{P} \sum _{I = 1} ^{P} \delta \( \mathbf{q} - \bar{\mathbf{q}} _{I} \( \mathbf{x}, t \) \) \approx \frac{1}{P} \sum _{I = 1} ^{P} K _{h} \( \mathbf{q} - \bar{\mathbf{q}} _{I} \( \mathbf{x}, t \) \).
  \label{eq:model_DeteParAppr}
\end{align}
So that the distribution of representative particles $\( \bar{\mathbf{q}} _{I} \) _{I \in \lbk 1, \cdots, P \rbk}$ contains the information of the polymer configuration density function $f$.
These representative particles are not real microscopic polymer molecules but coarse-grained particles.
Here, the smooth kernel function $K _{h}$ is usually chosen as the classical Gaussian kernel,
\begin{align*}
  K _{h} \( \mathbf{q} \) 
  = 
  \frac{1}{\[ \sqrt{2 \pi} h \] ^{\[ N - 1 \] \times d}} \exp \( - \frac{\lmdl \mathbf{q} \rmdl ^{2}}{2 h ^{2}} \),
\end{align*}
where $h \in \mathbb{R} _{+} \[ \mykgms{0}{1}{0} \]$ is the bandwidth of the kernel.
Following the strategy from \cite{bao2025deterministic}, we choose the bandwidth $h$ as
\begin{align}
  h = \frac{\textup{med}}{\sqrt{2 \log P}}
  ,
  \quad
  \textup{med} = \textup{median} _{J, K = 1, \cdots, P} \( \lmdl \bar{\mathbf{q}} _{J} - \bar{\mathbf{q}} _{K} \rmdl \).
  \label{eq:model_bandwidth}
\end{align}

Substituting \autoref{eq:model_DeteParAppr} into \autoref{eq:model_FP}, the evolution of the particle approximation $\bar{\mathbf{q}} _{k, I} \( \mathbf{x}, t \) \in \mathbb{R} ^{d} \[ \mykgms{0}{1}{0} \]$, for $I \in \lbk 1, \cdots, P \rbk$, $k \in \lbk 1, \cdots, N - 1 \rbk$, is \cite{bao2025micro}
\begin{align}
  &
  \quad
  \partial _{t} \bar{\mathbf{q}} _{k, I}
  +
  \mathbf{u} \cdot \nabla _{\mathbf{x}} \bar{\mathbf{q}} _{k, I}
  -
  \bar{\mathbf{q}} _{k, I} \cdot \nabla _{\mathbf{x}} \mathbf{u}
  =
  \dot{\bar{\mathbf{q}}} _{k, I}
  -
  \bar{\mathbf{q}} _{k, I} \cdot \nabla _{\mathbf{x}} \mathbf{u}
  \nonumber
  \\
  &
  =
  -
  \frac{1}{\zeta}
  \sum _{j = 1} ^{N - 1}
  A _{kj}
  \[ \kB T \[ \frac{\sum _{K} \nabla _{\mathbf{q} _{j}} K _{h} \( \bar{\mathbf{q}} _{I} - \bar{\mathbf{q}} _{K} \)}{\sum _{J} K _{h} \( \bar{\mathbf{q}} _{I} - \bar{\mathbf{q}} _{J} \)} + \sum _{K} \frac{\nabla _{\mathbf{q} _{j}} K _{h} \( \bar{\mathbf{q}} _{I} - \bar{\mathbf{q}} _{K} \)}{\sum _{J} K _{h} \( \bar{\mathbf{q}} _{J} - \bar{\mathbf{q}} _{K} \)} \] + \nabla _{\mathbf{q} _{j}} \Psi \( \bar{\mathbf{q}} _{I} \) \]
  .
  \label{eq:model_DymRepPar}
\end{align}
This is defined as the reference model.

Now, given any initial and boundary conditions of the position field $\( \bar{\mathbf{q}} _{I} \) _{I \in \lbk 1, \cdots, P \rbk}$, we are able to solve the dynamics of the representative particles through \autoref{eq:model_DymRepPar}, and then recover the dynamics of the polymer configuration density function $f$ through the deterministic particle approximation \autoref{eq:model_DeteParAppr}, which is the solution of the microscopic Fokker-Planck equation \autoref{eq:model_FP}.

Numerical experiments reveal that VDS solves the microscopic Fokker-Planck equation accurately in many different 2D dumbbell setups (even coupled with the Navier-Stokes equation), and costs less computational time compared with many other solvers \cite{bao2024micro, bao2025deterministic}. 

However, as we have mentioned in \autoref{sec:introduction}, the computational cost increases if we are working on 3D multi-bead cases, since we need more representative particles for accuracy.
To overcome this issue, we will discuss how to apply the MOR to accelerate the computation.


\subsection{POD-MOR with Shared Basis}

In this part, we set up the POD-MOR for the configuration of representative particles $\( \bar{\mathbf{q}} _{I} \) _{I \in \lbk 1, \cdots, P \rbk}$. 
Here, the reduced model contains $R \( \le Pd \)$ DoF, while the full model contains $F = \[ N - 1 \] P d$ DoF.

For the convenience of the notation in the following transformations, we will remove the upper bar $\bar{\cdot}$ from this point on.
We consider the configuration of representative particles as $\mathbf{q} \in \mathbb{R} ^{F} \[ \mykgms{0}{1}{0} \]$.

In order to construct the mapping between the full configuration and the reduced configuration, we define the shared POD matrix as
\begin{align*}
  \mathbf{U} \in \mathbb{R} ^{\[ P d \] \times R}.
\end{align*}
In fact, the columns of $\mathbf{U}$ are orthogonal to each other, and they are dominant modes informed from the snapshots of $\mathbf{q} _{k :}$, $k \in \lbk 1, \cdots, N - 1 \rbk$ during the evolution of the reference model. 
The calculation of $\mathbf{U}$ is detailed in \autoref{sec:experiments_setup}.

Now, we can calculate the reduced configuration by the map $\mathbf{p} _{k :} := \mathbf{U} \tp \cdot \mathbf{q} _{k :} \in \mathbb{R} ^{R} \[ \mykgms{0}{1}{0} \]$, and the approximated configuration $\tilde{\mathbf{q}} _{k :} := \mathbf{U} \cdot \mathbf{p} _{k :} \in \mathbb{R} ^{P d} \[ \mykgms{0}{1}{0} \]$.
The entry-wise expressions are given as,
\begin{align}
  p _{k s} = \sum _{\( I \alpha \) = 1} ^{P d} U _{\( I \alpha \) s} q _{k \( I \alpha \)},
  \quad
  \tilde{q} _{k \( I \alpha \)} = \sum _{s = 1} ^{R} U _{\( I \alpha \) s} p _{k s}.
  \label{eq:model_pq}
\end{align}
Since the same POD matrix $\mathbf{U}$ is applied for each bond $k \in \lbk 1, \cdots, N - 1 \rbk$, the reduced space is shared among all bonds.

In computation, we denote the velocity deformation gradient by $\mathbf{G} := \[ \nabla _{\mathbf{x}} \mathbf{u} \] \tp$ which is fixed.
Then, we rewrite the reference model \autoref{eq:model_DymRepPar}, i.e., the evolution of $\mathbf{q} _{k, I}$, for the particle index $I \in \lbk 1, \cdots, P \rbk$ and the bond index $k \in \lbk 1, \cdots, N - 1 \rbk$
\begin{align*}
  &
  \quad
  \partial _{t} \mathbf{q} _{k, I}
  +
  \mathbf{u} \cdot \nabla _{\mathbf{x}} \mathbf{q} _{k, I}
  -
  \mathbf{G} \cdot \mathbf{q} _{k, I}
  \nonumber
  \\
  &
  =
  -
  \frac{1}{\zeta}
  \sum _{j = 1} ^{N - 1}
  A _{kj}
  \[ \kB T \[ \frac{\sum _{K} \nabla _{\mathbf{q} _{j}} K _{h} \( \mathbf{q} _{I} - \mathbf{q} _{K} \)}{\sum _{J} K _{h} \( \mathbf{q} _{I} - \mathbf{q} _{J} \)} + \sum _{K} \frac{\nabla _{\mathbf{q} _{j}} K _{h} \( \mathbf{q} _{I} - \mathbf{q} _{K} \)}{\sum _{J} K _{h} \( \mathbf{q} _{J} - \mathbf{q} _{K} \)} \] + \nabla _{\mathbf{q} _{j}} \Psi \( \mathbf{q} _{I} \) \]
  .
\end{align*}

Using Galerkin projection, and given the reduced index $s \in \lbk 1, \cdots, R \rbk$ and the bond index $k \in \lbk 1, \cdots, N - 1 \rbk$, the POD-MOR model is given by
\begin{align}
  \sum _{I = 1} ^{P}
  \sum _{\alpha = 1} ^{d}
  U _{\( I \alpha \) s}
  \[
  \partial _{t} q _{k \( I \alpha \)}
  +
  \[ \mathbf{u} \cdot \nabla _{\mathbf{x}} \mathbf{q} _{k, I} \] _{\alpha}
  -
  \[ \mathbf{G} \cdot \mathbf{q} _{k, I} \] _{\alpha}
  \]
  &
  =
  -
  \frac{1}{\zeta}
  \sum _{I = 1} ^{P}
  \sum _{\alpha = 1} ^{d}
  \sum _{j = 1} ^{N - 1}
  A _{kj}
  U _{\( I \alpha \) s}
  S,
  \label{eq:model_DymRepParPod}
\end{align}
where
\begin{align}
  S
  :=
  \kB T \[ \frac{\sum _{K = 1} ^{P} \nabla _{j \alpha} K _{h} \( \mathbf{q} _{I} - \mathbf{q} _{K} \)}{\sum _{J = 1} ^{P} K _{h} \( \mathbf{q} _{I} - \mathbf{q} _{J} \)} + \sum _{K = 1} ^{P} \frac{\nabla _{j \alpha} K _{h} \( \mathbf{q} _{I} - \mathbf{q} _{K} \)}{\sum _{J = 1} ^{P} K _{h} \( \mathbf{q} _{J} - \mathbf{q} _{K} \)} \] + \nabla _{j \alpha} \Psi \( \mathbf{q} _{I} \)
  .
\end{align}

\subsection{Acceleration of POD-MOR}

In order to achieve a reduced evolution of $\mathbf{p}$, now we write each term in \autoref{eq:model_DymRepParPod} as a function of reduced coordinate $\mathbf{p}$ by approximating some of $\mathbf{q} _{k :}$ by $\tilde{\mathbf{q}} _{k :} = \mathbf{U} \cdot \mathbf{p} _{k :}$.

Given the reduced index $s \in \lbk 1, \cdots, R \rbk$ and the bond index $k \in \lbk 1, \cdots, N - 1 \rbk$, the terms on the LHS of \autoref{eq:model_DymRepParPod} are listed as following.
\begin{itemize}
  \item The time derivative term is linear
    \begin{align*}
      \sum _{I = 1} ^{P}
      \sum _{\alpha = 1} ^{d}
      U _{\( I \alpha \) s}
      \partial _{t} q _{k \( I \alpha \)}
      \approx
      \sum _{I = 1} ^{P}
      \sum _{\alpha = 1} ^{d}
      U _{\( I \alpha \) s}
      \partial _{t}
      \sum _{t = 1} ^{R}
      U _{\( I \alpha \) t} p _{k t}
      =
      \partial _{t}
      p _{k s}
      .
    \end{align*}
  \item The convection term is linear
    \begin{align*}
      &
      \quad
      \sum _{I = 1} ^{P}
      \sum _{\alpha = 1} ^{d}
      U _{\( I \alpha \) s}
      \[ \mathbf{u} \cdot \nabla _{\mathbf{x}} \mathbf{q} _{k, I} \] _{\alpha}
      =
      \sum _{I = 1} ^{P}
      \sum _{\alpha, \beta = 1} ^{d}
      U _{\( I \alpha \) s}
      u _{\beta} \nabla _{x _{\beta}} q _{k \( I \alpha \)}
      \\
      &
      \approx
      \sum _{I = 1} ^{P}
      \sum _{\alpha, \beta = 1} ^{d}
      U _{\( I \alpha \) s}
      u _{\beta} \nabla _{x _{\beta}}
      \sum _{t = 1} ^{R}
      U _{\( I \alpha \) t} p _{k t}
      =
      \sum _{\beta = 1} ^{d}
      u _{\beta} 
      \nabla _{x _{\beta}} p _{k s}
      .
    \end{align*}
  \item The hydrodynamic coupling term is linear
    \begin{align*}
      &
      \quad
      \sum _{I = 1} ^{P}
      \sum _{\alpha = 1} ^{d}
      U _{\( I \alpha \) s}
      \[ \mathbf{G} \cdot \mathbf{q} _{k, I} \] _{\alpha}
      =
      \sum _{I = 1} ^{P}
      \sum _{\alpha, \beta = 1} ^{d}
      U _{\( I \alpha \) s}
      G _{\alpha \beta} q _{k \( I \beta \)}
      \\
      &
      \approx
      \sum _{I = 1} ^{P}
      \sum _{\alpha, \beta = 1} ^{d}
      U _{\( I \alpha \) s}
      G _{\alpha \beta}
      \sum _{t = 1} ^{R}
      U _{\( I \beta \) t} p _{k t}
      =:
      \sum _{t = 1} ^{R}
      X _{st} ^{H} p _{k t}
      ,
    \end{align*}
    where
    \begin{align*}
      X _{st} ^{H} 
      :=
      \sum _{I = 1} ^{P}
      \sum _{\alpha, \beta = 1} ^{d}
      U _{\( I \alpha \) s}
      G _{\alpha \beta}
      U _{\( I \beta \) t}
      .
    \end{align*}
\end{itemize}

Note that the LHS of \autoref{eq:model_DymRepParPod} in the reduced coordinate is linear. 
However, some terms on the RHS of \autoref{eq:model_DymRepParPod} are nonlinear, so directly solving the evolution in the reduced coordinate $\mathbf{p}$ requires solving $\mathbf{q}$ and then applying the projection. The issue is that this step requires more computational time than that of the original model.
To reduce the computational cost, we need to linearize the terms on the RHS as following.
\begin{itemize}
  \item The intra-molecule interaction term is linear when choosing the Hookean potential $\Psi \( \mathbf{q} _{I} \) = \sum _{j = 1} ^{N - 1} \sum _{\alpha = 1} ^{d} H _{j} q _{j \( I \alpha \)} ^{2} / 2$,
    \begin{align*}
      &
      \quad 
      \sum _{I = 1} ^{P}
      \sum _{\alpha = 1} ^{d}
      \sum _{j = 1} ^{N - 1}
      A _{kj}
      U _{\( I \alpha \) s}
      \nabla _{j \alpha} \Psi \( \mathbf{q} _{I} \)
      =
      \sum _{I = 1} ^{P}
      \sum _{\alpha = 1} ^{d}
      \sum _{j = 1} ^{N - 1}
      H _{j}
      A _{kj}
      U _{\( I \alpha \) s}
      q _{j \( I \alpha \)}
      \\
      &
      \approx
      \sum _{I = 1} ^{P}
      \sum _{\alpha = 1} ^{d}
      \sum _{j = 1} ^{N - 1}
      H _{j}
      A _{kj}
      U _{\( I \alpha \) s}
      \sum _{t = 1} ^{R}
      U _{\( I \alpha \) t} p _{j t}
      =
      \sum _{j = 1} ^{N - 1}
      H _{j}
      A _{kj}
      p _{j s}
      .
    \end{align*}
  \item Both of the terms in the Brownian motion with a Gaussian kernel are nonlinear, we linearize the first term as
    \begin{align*}
      &
      \quad 
      \sum _{I = 1} ^{P}
      \sum _{\alpha = 1} ^{d}
      \sum _{j = 1} ^{N - 1}
      A _{kj}
      U _{\( I \alpha \) s}
      \frac{\sum _{K = 1} ^{P} \nabla _{j \alpha} K _{h} \( \mathbf{q} _{I} - \mathbf{q} _{K} \)}{\sum _{J = 1} ^{P} K _{h} \( \mathbf{q} _{I} - \mathbf{q} _{J} \)}
      \nonumber
      \\
      &
      =
      - \frac{1}{h ^{2}}
      \sum _{I = 1} ^{P}
      \sum _{\alpha = 1} ^{d}
      \sum _{j = 1} ^{N - 1}
      A _{kj}
      U _{\( I \alpha \) s}
      \frac{\sum _{K = 1} ^{P} K _{h} \( \mathbf{q} _{I} - \mathbf{q} _{K} \) \[ q _{j \( I \alpha \)} - q _{j \( K \alpha \)} \]}{\sum _{J = 1} ^{P} K _{h} \( \mathbf{q} _{I} - \mathbf{q} _{J} \)}
      \\
      &
      \approx
      - \frac{1}{h ^{2}}
      \sum _{I = 1} ^{P}
      \sum _{\alpha = 1} ^{d}
      \sum _{j = 1} ^{N - 1}
      A _{kj}
      U _{\( I \alpha \) s}
      \frac{\sum _{K = 1} ^{P} K _{h} \( \mathbf{q} _{I} - \mathbf{q} _{K} \) \sum _{t = 1} ^{R} \[ U _{\( I \alpha \) t} - U _{\( K \alpha \) t} \] p _{j t}}{\sum _{J = 1} ^{P} K _{h} \( \mathbf{q} _{I} - \mathbf{q} _{J} \)}
      \\
      &
      =:
      - \frac{1}{h ^{2}}
      \sum _{I = 1} ^{P}
      \sum _{\alpha = 1} ^{d}
      U _{\( I \alpha \) s}
      \sum _{t = 1} ^{R}
      \frac{\sum _{K = 1} ^{P} Z _{IK} Y _{IK \alpha t}}{\sum _{J = 1} ^{P} Z _{IJ}}
      \sum _{j = 1} ^{N - 1}
      A _{kj} p _{j t}
      \\
      &
      =:
      \sum _{j = 1} ^{N - 1}
      A _{kj}
      \sum _{t = 1} ^{R}
      X _{st} ^{B1} p _{j t}
      .
    \end{align*}
  \item We linearize the second term of the Brownian motion as
    \begin{align*}
      &
      \quad 
      \sum _{I = 1} ^{P}
      \sum _{\alpha = 1} ^{d}
      \sum _{j = 1} ^{N - 1}
      A _{kj}
      U _{\( I \alpha \) s}
      \sum _{K = 1} ^{P} \frac{\nabla _{j \alpha} K _{h} \( \mathbf{q} _{I} - \mathbf{q} _{K} \)}{\sum _{J = 1} ^{P} K _{h} \( \mathbf{q} _{J} - \mathbf{q} _{K} \)}
      \nonumber
      \\
      &
      =
      - \frac{1}{h ^{2}}
      \sum _{I = 1} ^{P}
      \sum _{\alpha = 1} ^{d}
      \sum _{j = 1} ^{N - 1}
      A _{kj}
      U _{\( I \alpha \) s}
      \sum _{K = 1} ^{P} \frac{K _{h} \( \mathbf{q} _{I} - \mathbf{q} _{K} \) \[ q _{j \( I \alpha \)} - q _{j \( K \alpha \)} \]}{\sum _{J = 1} ^{P} K _{h} \( \mathbf{q} _{J} - \mathbf{q} _{K} \)}
      \\
      &
      \approx
      - \frac{1}{h ^{2}}
      \sum _{I = 1} ^{P}
      \sum _{\alpha = 1} ^{d}
      \sum _{j = 1} ^{N - 1}
      A _{kj}
      U _{\( I \alpha \) s}
      \sum _{K = 1} ^{P} \frac{K _{h} \( \mathbf{q} _{I} - \mathbf{q} _{K} \) \sum _{t = 1} ^{R} \[ U _{\( I \alpha \) t} - U _{\( K \alpha \) t} \] p _{j t}}{\sum _{J = 1} ^{P} K _{h} \( \mathbf{q} _{J} - \mathbf{q} _{K} \)}
      \\
      &
      =:
      - \frac{1}{h ^{2}}
      \sum _{I = 1} ^{P}
      \sum _{\alpha = 1} ^{d}
      U _{\( I \alpha \) s}
      \sum _{t = 1} ^{R}
      \sum _{K = 1} ^{P}
      \frac{Z _{IK} Y _{IK \alpha t}}{\sum _{J = 1} ^{P} Z _{JK}}
      \sum _{j = 1} ^{N - 1}
      A _{kj} p _{j t}
      \\
      &
      =:
      \sum _{j = 1} ^{N - 1}
      A _{kj}
      \sum _{t = 1} ^{R}
      X _{st} ^{B2} p _{j t}
      .
    \end{align*}
\end{itemize}
Again, $X _{st} ^{B1}$ and $X _{st} ^{B2}$ denote the coefficients shown above.

Following the above approximation, we obtain the evaluation equations for the reduced system for any reduced index $s \in \lbk 1, \cdots, R \rbk$ and any bond index $k \in \lbk 1, \cdots, N - 1 \rbk$,
\begin{align}
  &
  \quad
  \partial _{t}
  p _{k s}
  +
  \sum _{\beta = 1} ^{d}
  u _{\beta} 
  \nabla _{x _{\beta}} p _{k s}
  -
  \sum _{t = 1} ^{R}
  X _{st} ^{H} p _{k t}
  \nonumber
  \\
  &
  =
  -\frac{1}{\zeta} 
  \[ 
  \kB T
  \sum _{j = 1} ^{N - 1}
  A _{kj}
  \sum _{t = 1} ^{R}
  \[ X _{st} ^{B1} + X _{st} ^{B2} \] p _{j t}
  +
  \sum _{j = 1} ^{N - 1}
  H _{j}
  A _{kj}
  p _{j s} 
  \]
  .
  \label{eq:model_DymReduced}
\end{align}

Since the DoF of the system \autoref{eq:model_DymReduced} is $\[ N - 1 \] R$ can be much smaller than the $F = \[ N - 1 \] P d$, it is possible to accelerate the computation.
However, a direct simulation of \autoref{eq:model_DymReduced} usually results in a large modeling error due to the nonlinearity of \autoref{eq:model_DymRepPar}. Therefore, we will correct the integration error by mapping back and forth using \autoref{eq:model_pq} after certain number of steps.

In the next section, we will implement the POD-MOR with VDS in some specific cases and check the acceleration and the MOR error.

\section{Numerical Experiments}
\label{sec:experiments}

This section presents a comprehensive set of numerical experiments designed to validate the performance and robustness of the proposed POD-MOR framework of VDS for 3D multi-bead polymer molecules by the comparison between the reduced model and the original model, in the sense of the computational time and the reduction error.

\subsection{Model Setup}
\label{sec:experiments_setup}

In VDS, we focus on the evolution of representative particles $\mathbf{q}$.
All representative particles are initialized from the thermal equilibrium distribution. 
At thermal equilibrium, the polymer configuration density function $f = f _{\textup{eq}}$ satisfies \autoref{eq:model_FP} with zero LHS, which implies the Boltzmann distribution
\begin{align*}
  f _{\textup{eq}} \( \mathbf{q} \) \sim \exp \( - \frac{\Psi \( \mathbf{q} \)}{\kB T} \).
\end{align*}
In order to generate initial representative particles, the components are generated from the normal distribution, for all bond $k \in \lbk 1, \cdots, N - 1 \rbk$ and spatial dimension $\alpha \in \lbk 1, \cdots, d \rbk$,
\begin{align*}
  q _{k \( I \alpha \)} \sim \mathcal{N} \( 0, \frac{\kB T}{H _{k}} \).
\end{align*}
At equilibrium, the theoretical value \cite{bird1987dynamics} of the polymer stress $\boldsymbol{\tau}$ should be 
\begin{align}
  \boldsymbol{\tau} = n \[ N - 1 \] \kB T \Id,
  \label{eq:experiments_equilibriumStress}
\end{align}
where $\Id \in \mathbb{R} ^{d \times d}$ is the identity matrix. Here we fix the polymer number density as $n = 1 \[ \mykgms{0}{-d}{0} \]$ and the absolute temperature $T = 1 \[ \textup{K} \]$.

We study the two different cases:
\begin{enumerate}
  \item no flow, the velocity gradient $\nabla _{\mathbf{x}} \mathbf{u} = \mathbf{0}$, i.e., relaxation process;
  \item simple shear flow, the velocity gradient $\nabla _{\mathbf{x}} \mathbf{u}$ has only one nonzero element $\nabla _{x _{1}} u _{2} = 1$.
\end{enumerate}
In both cases, we can assume the configuration of representative particles $\mathbf{q} \( \mathbf{x}, t \)$ is homogeneous in the flow direction, i.e., the convection term in \autoref{eq:model_DymRepPar} vanishes
\begin{align*}
  \mathbf{u} \cdot \nabla _{\mathbf{x}} \mathbf{q} = \mathbf{0}.
\end{align*}
We solve the reference model \autoref{eq:model_DymRepPar} and the POD-MOR model \autoref{eq:model_DymRepParPod} on a single spatial point $\mathbf{x}$, and the evolution is just ODEs with respect to time $t$.

We fix the friction coefficient ($\zeta = 4 \[ \mykgms{1}{0}{-1} \]$), and apply Hookean potentials \autoref{eq:model_Hookean} on molecule bonds. Both cases with homogeneous and inhomogeneous bond coefficients are studied.
For the dumbbell cases with simple shear flow, we set the Weissenberg number $\textup{Wi} = 1$, where it measures the quotient between the typical velocity and the typical length as the velocity gradient,
\begin{align*}
  \textup{Wi} = \lnm \nabla _{\mathbf{x}} \mathbf{u} \rnm \zeta / 4 H,
\end{align*}
so that $H = 1$.
For homogeneous multi-bead cases, we fix $H _{k} \equiv 1$. 
For inhomogeneous multi-bead cases, we take $H _{k} = k$, $k \in \lbk 1, \cdots, N - 1 \rbk$.

We use the explicit Euler method for time integration in both models \autoref{eq:model_DymRepPar} and \autoref{eq:model_DymRepParPod} with time step size $\delta _{\textup{t}}$.
Since in the original references \cite{bao2024micro, bao2025deterministic}, the reference model is only implemented with 2D dumbbell polymer molecules, and the Fokker-Planck equation is coupled with the Navier-Stokes equation using a time-implicit scheme, for a complete study, we need to verify the accuracy of the current reference model \autoref{eq:model_DymRepPar} in \autoref{sec:accuracy} by the convergence test with respect to parameters in the numerical scheme. 

For both no flow and the simple shear flow cases, we investigate the predictive capability from simulations over a relative long time period. Specifically, we run a reference dynamics \autoref{eq:model_DymRepPar} from $t = 0$ to $t = 6$, and save the trajectory as $\mathbf{q} ^{\textup{ref}} \( t \)$. Then we collect POD data from $t = 0$ to $t = 3$ to calculate the POD matrix $\mathbf{U}$. After that, we run the POD-MOR dynamics \autoref{eq:model_DymRepParPod} from $t = 0$ to $t = 6$ with the same model parameters and then recover the dynamics of the full representative particles $\mathbf{q} ^{\textup{mor}} \( t \)$ using \autoref{eq:model_pq}.

The relative $L ^{2}$ MOR error can be calculated for any time $t$,
\begin{align*}
  \textup{err} _{\textup{MOR}} \( t \)
  =
  \frac{\lnm \mathbf{q} ^{\textup{mor}} \( t \) - \mathbf{q} ^{\textup{ref}} \( t \) \rnm _{L ^{2}}}{\lnm \mathbf{q} ^{\textup{ref}} \( t \) \rnm _{L ^{2}}}
  .
\end{align*}
Since the POD-MOR is applied to a Lagrangian system, the particle index matters, and the same initial condition is used in both the reference dynamics and the MOR dynamics ($\mathbf{q} ^{\textup{ref}} \( 0 \) = \mathbf{q} ^{\textup{mor}} \( 0 \)$).

Rather than finding dominant modes by solving eigenvalue problems, the method of snapshots provides an alternate way of computing the POD modes \cite{rowley2004model}.
To collect the POD data, we first collect $L + 1$ snapshots $\( \mathbf{q} ^{\textup{ref}} \( t _{l} \) \) _{l \in \lbk 0, \cdots, L \rbk}$ at the uniform time spacing, where $0 = t _{0} < t _{1} < \cdots < t _{L} = 3$. The POD data contains all bond segments, i.e., each snapshot generates $N - 1$ POD data. The whole set is denoted as $\mathbf{X}$, where
\begin{align*}
  \mathbf{X} = \( \mathbf{q} _{k} ^{\textup{ref}} \( t _{l} \) \) _{l \in \lbk 0, \cdots, L \rbk, k \in \lbk 1, \cdots, N - 1 \rbk} \in \mathbb{R} ^{Pd \times \[ L + 1 \] \[ N - 1 \]}.
\end{align*}
We derive the POD matrix $\mathbf{U}$ by applying singular value decomposition (SVD) to $\mathbf{X} \cdot \mathbf{X} \tp$,
\begin{align*}
  \mathbf{X} \cdot \mathbf{X} \tp = \hat{\mathbf{U}} \cdot \hat{\boldsymbol{\Sigma}} \cdot \hat{\mathbf{U}} \tp,
\end{align*}
and we choose $\mathbf{U}$ as the first $R$ columns of $\hat{\mathbf{U}}$ corresponding to the $R$ dominant singular values in the diagonal matrix $\hat{\boldsymbol{\Sigma}}$, where $R$ is the reduced DoF. 
The above decomposition shows that POD automatically identifies the most dominant spatial modes so that it will provide deeper insights into the underlying physics.

In numerical experiments, we choose 
\begin{align}
  L = \frac{Pd}{N - 1} = \frac{F}{ \[ N - 1 \] ^{2}}
  \label{eq:experiments_snapshot}
\end{align}
to balance the computation and the expressibility (the column space of $\hat{\mathbf{U}}$), and we pick a small reduced DoF $R \ll L$.

The numerical experiments are run on i7-13700k CPU with 16 cores, 3.4GHz to 5.3GHz Turbo. We denote $T ^{\textup{ref}}$ as the computational time of the temporal iteration part for the reference dynamics, and denote $T ^{\textup{mor}}$ as the computational time of the online temporal iteration part for the reduced (POD-MOR) dynamics.

\subsection{Predictive Capability for Long-term Simulations --- No Flow}

Here we test both models \autoref{eq:model_DymRepPar} and \autoref{eq:model_DymRepParPod} without velocity terms.

We fix the number of representative particles $P = 1000$, time step size $\delta _{\textup{t}} = 0.001$, and freeze nonlinear terms every $300$ step.

For the system of $2$-bead (dumbbell) polymer molecules, we choose a number of snapshots $L = 3000$, and for the system of $3$-bead polymer molecules, we choose a number of snapshots $L = 1500$.
Such a choice is based on \autoref{eq:experiments_snapshot}.

\begin{figure}[!ht]
  \centering
  \includegraphics[width = 0.5\textwidth]{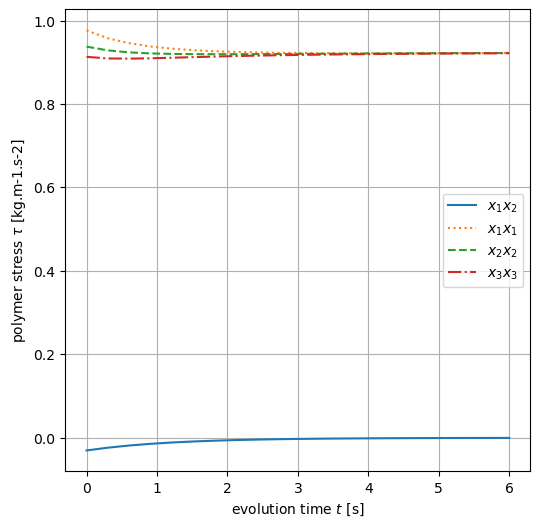}
  \caption{Polymer stress $\boldsymbol{\tau}$ vs. evolution time $t$ in the reference dynamics. Parameters: $2$-bead polymer molecules without flow, particle number $P = 1000$, time step size $\delta _{t} = 0.001$.}
  \label{fig:experiments_noflow_2bead_stress}
\end{figure}

\begin{figure}[!ht]
  \centering
  \includegraphics[width = 0.24\textwidth]{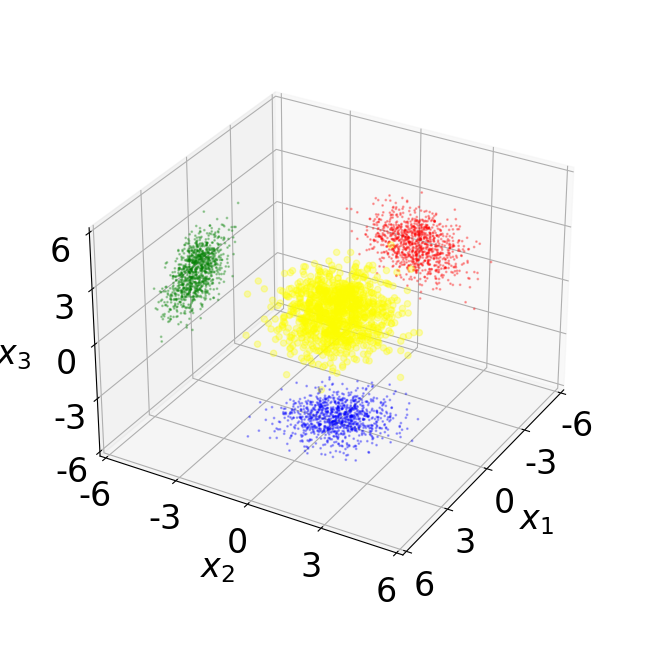}
  \includegraphics[width = 0.24\textwidth]{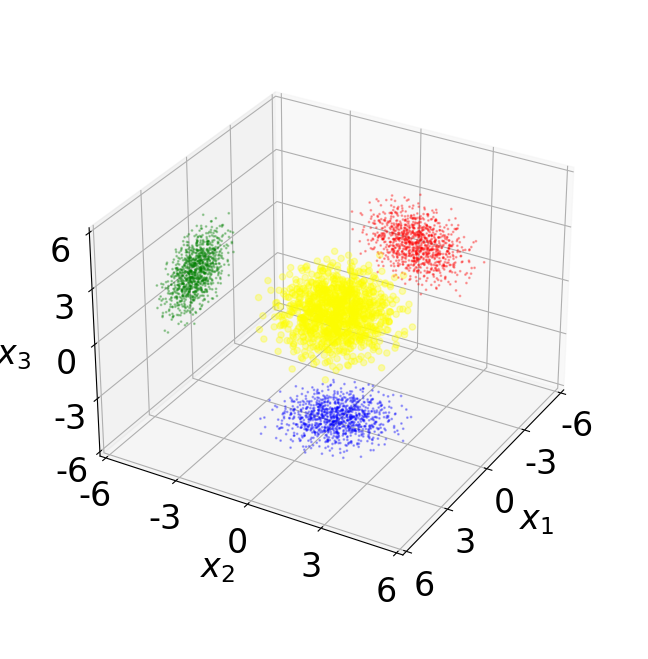}
  \includegraphics[width = 0.24\textwidth]{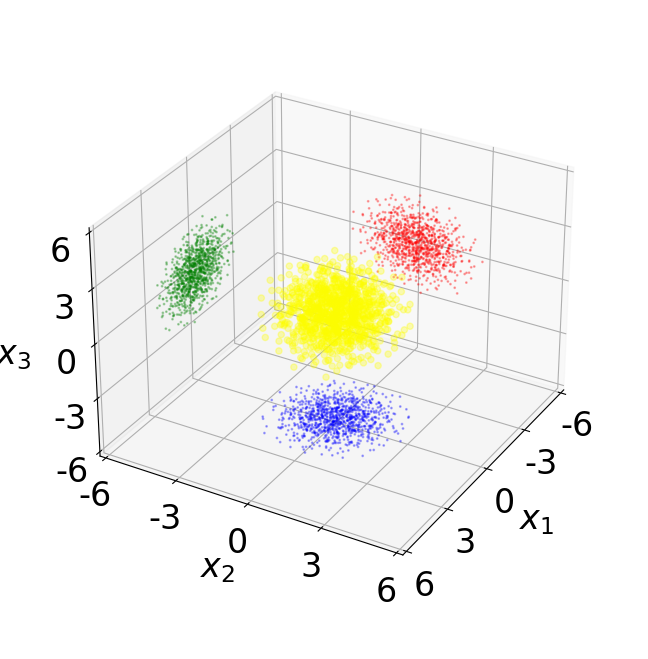}
  \includegraphics[width = 0.24\textwidth]{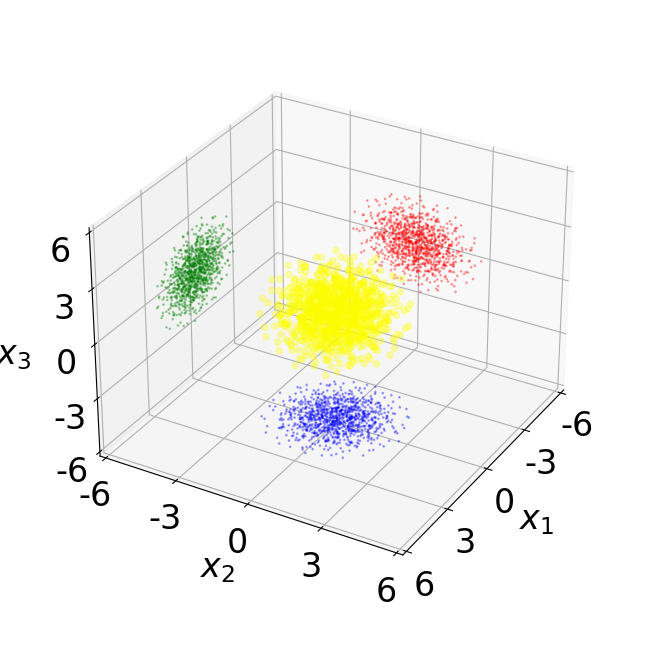}
  \caption{The bond of representative particles $\mathbf{q} _{1:}$ in the reference dynamics. From left to right, the plots correspond to evolution time $t = 0, 1.5, 3, 6 \[ \textup{s} \]$. The bonds are illustrated in 3D space (yellow) and 2D projections on $x _{1} x _{2}$ (blue), $x _{2} x _{3}$ (red), $x _{3} x _{1}$ (green) planes. Parameters: $2$-bead polymer molecules without flow, particle number $P = 1000$, time step size $\delta _{t} = 0.001$.}
  \label{fig:experiments_noflow_2bead_particle}
\end{figure}

The reference dynamics \autoref{eq:model_DymRepPar} shows a steady evolution, and we plot the polymer stress in \autoref{fig:experiments_noflow_2bead_stress} and the evolution of the representative particles in \autoref{fig:experiments_noflow_2bead_particle} for the system of $2$-bead (dumbbell) polymer molecules.


When the representative particles take their initial values from the thermal equilibrium distribution, compared with the theoretical value of the polymer stress \autoref{eq:experiments_equilibriumStress}, the numerical evolution results in an underestimation on the diagonal entries of the polymer stress $\boldsymbol{\tau}$, with about $8\%$ relative error (see \autoref{fig:experiments_noflow_2bead_stress}). Since we focus on the POD-ROM of the VDS in this paper, we will use this dynamics as our reference solution. 

Looking at the different snapshots of the representative particles during the evolution in \autoref{fig:experiments_noflow_2bead_particle}, the distribution seems unchanged. Actually, the particles slow down and converge to the steady state quickly.

We take $R = 1, 2, 3, 4, 5, 10, 20, 40$, and we list the computational time and MOR error in \autoref{tab:experiments_noflow_2Bead} and \autoref{tab:experiments_noflow_3Bead} for $2$-bead and $3$-bead, respectively.

\begin{table}[!ht]
  \centering
  \begin{tabular}{|c|c|c|c|c|c|c|c|c|}
    \hline
    \multicolumn{9}{|c|}{Full model: $T ^{\textup{ref}} = 394$ (seconds), $F = 3000$.}
    \\
    \hline
    \hline
    $R$ & $1$ & $2$ & $3$ & $4$ & $5$ & $10$ & $20$ & $40$
    \\
    \hline
    $T ^{\textup{mor}} / T ^{\textup{ref}}$ & $0.360\%$ & $0.605\%$ & $0.819\%$ & $1.08\%$ & $1.43\%$ & $3.66\%$ & $11.8\%$ & $43.3\%$
    \\
    \hline
    $\textup{err} _{\textup{MOR}} \( t = 6 \)$ & $4.76\%$ & $2.34\%$ & $1.04\%$ & $0.635\%$ & $0.595\%$ & $0.816\%$ & $0.843\%$ & $0.865\%$
    \\
    \hline
  \end{tabular}
  \caption{The relation among the reduced DoF $R$, the ratio of computational time $T ^{\textup{mor}} / T ^{\textup{ref}}$, and the relative $L ^{2}$ MOR error $\textup{err} _{\textup{MOR}} \( t = 6 \)$, for different reduced DoF $R$ in the POD-MOR dynamics. Parameters: $2$-bead polymer molecules without flow, particle number $P = 1000$, time step size $\delta _{t} = 0.001$, number of snapshots $L = 3000$.}
  \label{tab:experiments_noflow_2Bead}
\end{table}

\begin{table}[!ht]
  \centering
  \begin{tabular}{|c|c|c|c|c|c|c|c|c|}
    \hline
    \multicolumn{9}{|c|}{Full model: $T ^{\textup{ref}} = 533$ (seconds), $F = 6000$.}
    \\
    \hline
    \hline
    $R$ & $1$ & $2$ & $3$ & $4$ & $5$ & $10$ & $20$ & $40$
    \\
    \hline
    $T ^{\textup{mor}} / T ^{\textup{ref}}$ & $0.441\%$ & $0.612\%$ & $0.758\%$ & $0.955\%$ & $1.23\%$ & $2.82\%$ & $8.90\%$ & $32.1\%$
    \\
    \hline
    $\textup{err} _{\textup{MOR}} \( t = 6 \)$ & $70.8\%$ & $12.8\%$ & $7.50\%$ & $4.47\%$ & $2.72\%$ & $1.82\%$ & $1.98\%$ & $1.98\%$
    \\
    \hline
  \end{tabular}
  \caption{The relation among the reduced DoF $R$, the ratio of computational time $T ^{\textup{mor}} / T ^{\textup{ref}}$, and the relative $L ^{2}$ MOR error $\textup{err} _{\textup{MOR}} \( t = 6 \)$, for different reduced DoF $R$ in the POD-MOR dynamics. Parameters: $3$-bead (homogeneous bonds) polymer molecules without flow, particle number $P = 1000$, time step size $\delta _{t} = 0.001$, number of snapshots $L = 1500$.}
  \label{tab:experiments_noflow_3Bead}
\end{table}

The results indicate that even if we use a single POD mode ($R = 1$) for $2$-bead, the relative error is less than $5\%$, and the same for $3$-bead when $R = 4$. That is reasonable since the initial and final distributions are similar.

\subsection{Predictive Capability for Long-term Simulations --- Simple Shear Flow}

Here we test both models \autoref{eq:model_DymRepPar} and \autoref{eq:model_DymRepParPod} with a given simple shear flow, where the velocity gradient $\nabla _{\mathbf{x}} \mathbf{u}$ has only one nonzero element $\nabla _{x _{1}} u _{2} = 1$.

We fix the number of representative particles $P = 1000$ for $2$-bead and $3$-bead polymer molecules, and choose $P = 1000, 2000, 5000, 10000$ for $4$-bead polymer molecules. We fix time step size $\delta _{\textup{t}} = 0.001$, and freeze nonlinear terms every $300$ step.
The setting is the same as the no-flow case.

For the system of $2$-bead (dumbbell) polymer molecules, we choose the number of snapshots $L = 3000$, for the system of $3$-bead polymer molecules, we choose the number of snapshots $L = 1500$, and for the system of $4$-bead polymer molecules, we choose the number of snapshots $L = 1000$, respectively, based on \autoref{eq:experiments_snapshot}. Additionally, we also consider the number of snapshots $L = 2000, 5000, 10000$ for $4$-bead polymer molecules for comparison on different number of representative particles.

\begin{figure}[!ht]
  \centering
  \includegraphics[width = 0.5\textwidth]{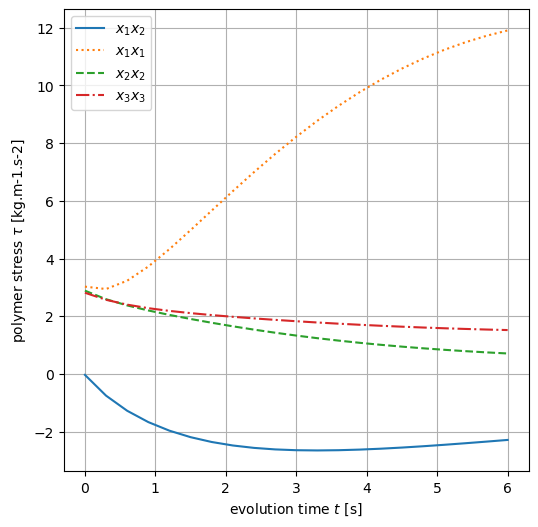}
  \caption{Polymer stress $\boldsymbol{\tau}$ vs. evolution time $t$ in the reference dynamics. Parameters: $4$-bead (homogeneous bonds) polymer molecules with simple shear flow, particle number $P = 1000$, time step size $\delta _{t} = 0.001$.}
  \label{fig:experiments_pureflow_4bead_stress}
\end{figure}

\begin{figure}[!ht]
  \centering
  \includegraphics[width = 0.24\textwidth]{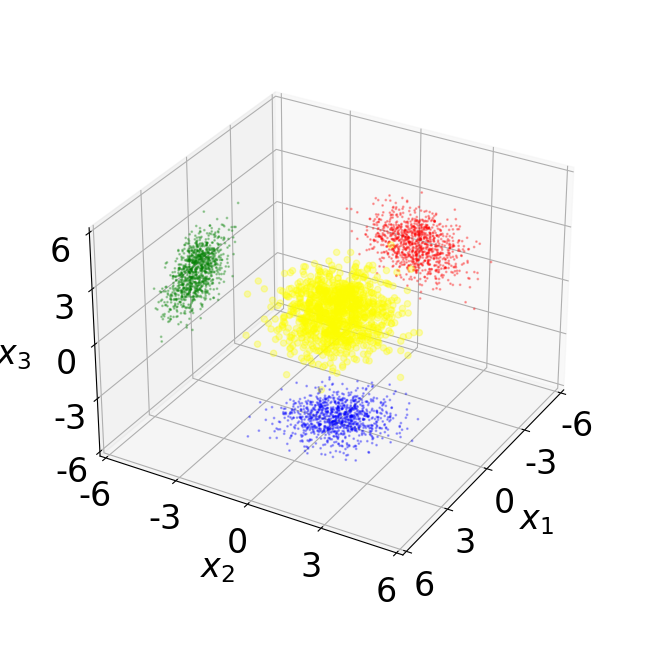}
  \includegraphics[width = 0.24\textwidth]{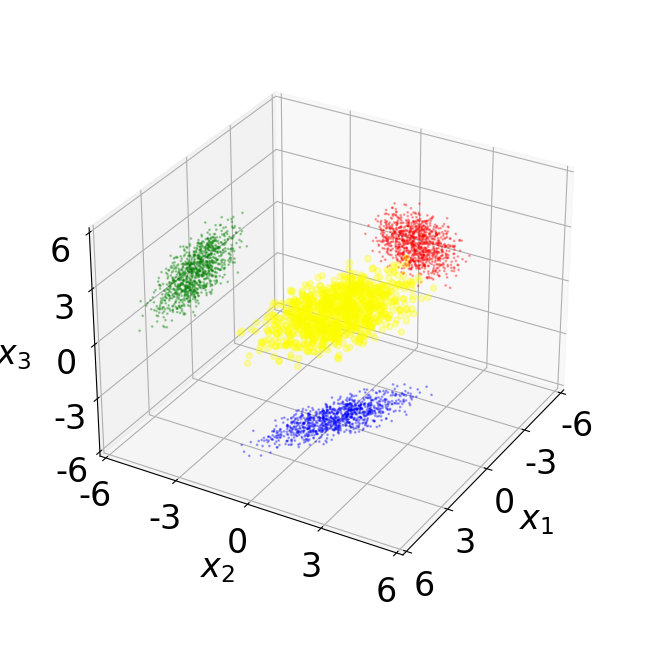}
  \includegraphics[width = 0.24\textwidth]{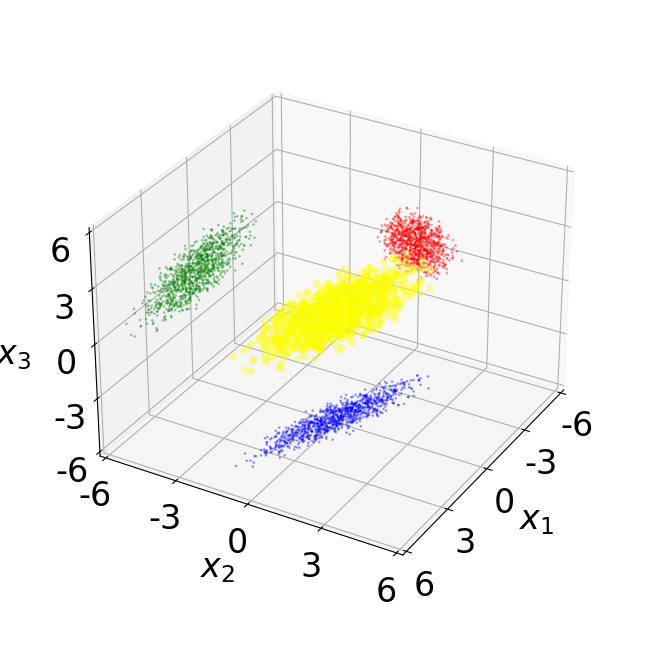}
  \includegraphics[width = 0.24\textwidth]{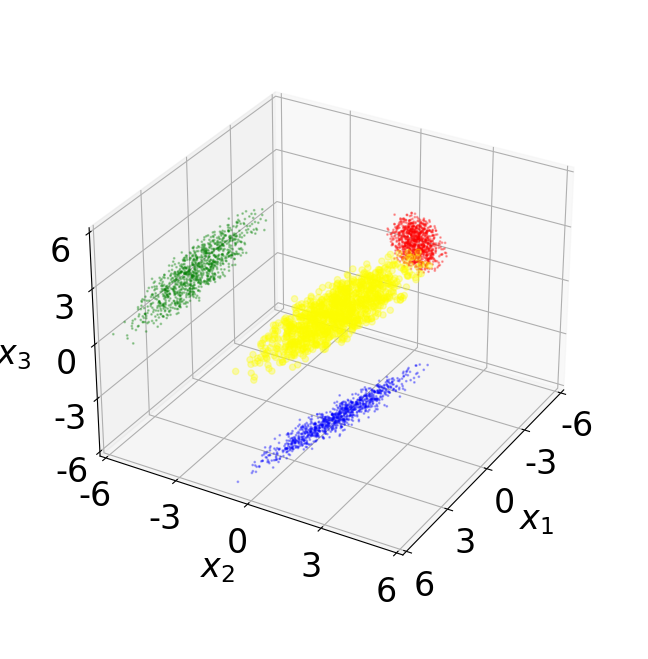}
  \\
  \includegraphics[width = 0.24\textwidth]{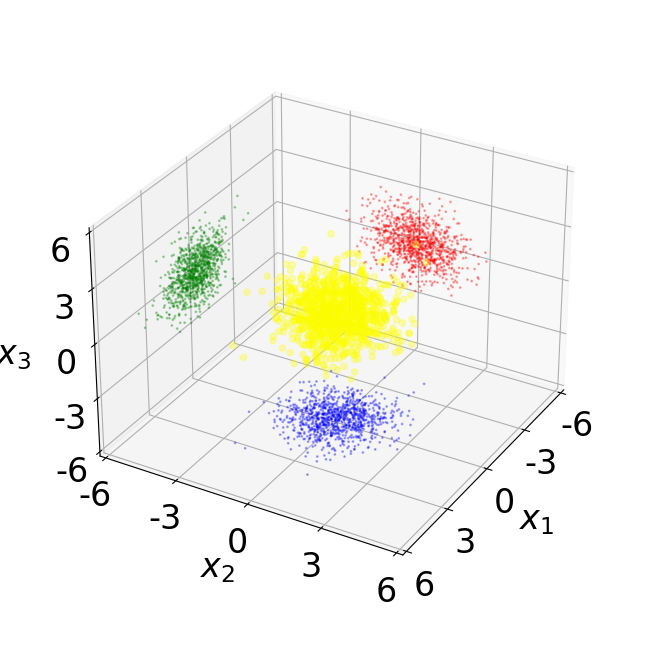}
  \includegraphics[width = 0.24\textwidth]{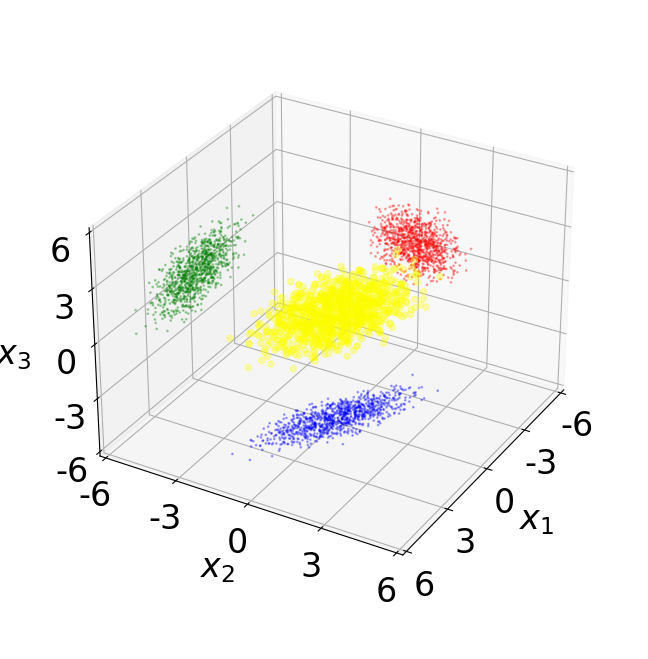}
  \includegraphics[width = 0.24\textwidth]{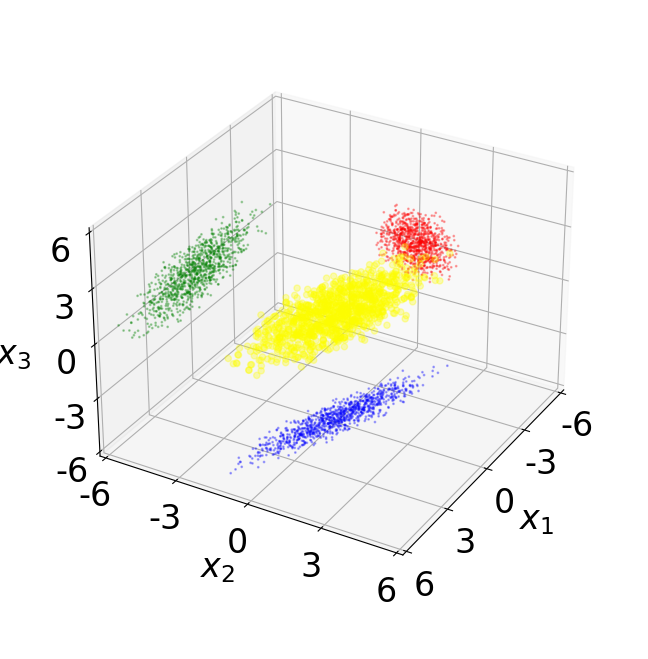}
  \includegraphics[width = 0.24\textwidth]{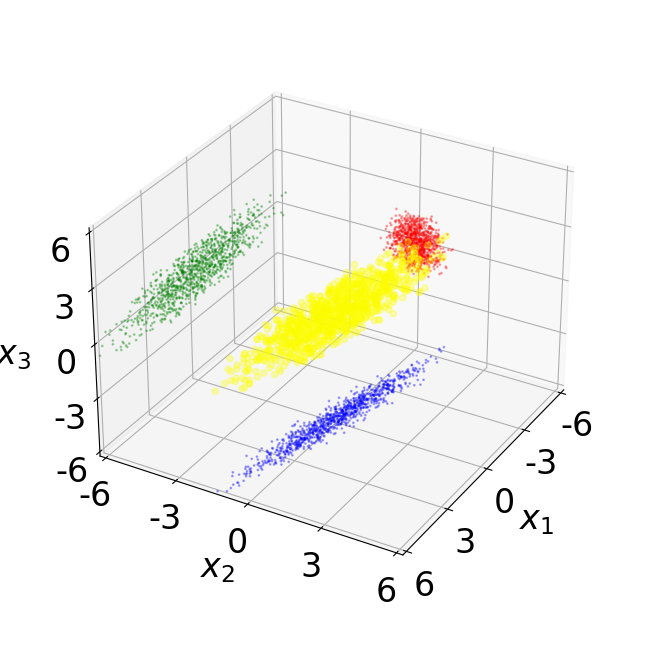}
  \caption{The bond of representative particles $\mathbf{q} _{1:}$ and $\mathbf{q} _{2:}$ in the reference dynamics. From top to bottom, the plots correspond to the first bond $\mathbf{q} _{1:}$ and the second bond $\mathbf{q} _{2:}$. From left to right, the plots correspond to evolution time $t = 0, 1.5, 3, 6 \[ \textup{s} \]$. The bonds are illustrated in 3D space (yellow) and 2D projections on $x _{1} x _{2}$ (blue), $x _{2} x _{3}$ (red), $x _{3} x _{1}$ (green) planes. Parameters: $4$-bead (homogeneous bonds) polymer molecules with simple shear flow, particle number $P = 1000$, time step size $\delta _{t} = 0.001$.}
  \label{fig:experiments_pureflow_4bead_particle}
\end{figure}

We plot the polymer stress in \autoref{fig:experiments_pureflow_4bead_stress} and the evolution of the representative particles in \autoref{fig:experiments_pureflow_4bead_particle} for the system of $4$-bead polymer molecules with $P = 1000$ representative particles.


Since the only nonzero element in the velocity gradient matrix is the term $\nabla _{x _{1}} u _{2}$, the polymer molecules tend to be parallel to the $x _{1}$ direction during the evolution due to the hydrodynamic effect, which can be observed from \autoref{fig:experiments_pureflow_4bead_particle}. We also notice that the bonds $\mathbf{q} _{1:}$ and $\mathbf{q} _{2:}$ reach different final distributions, and the middle bond $\mathbf{q} _{2:}$ is stretched more than the side bond $\mathbf{q} _{1:}$. As the bonds $\mathbf{q} _{1:}$ and $\mathbf{q} _{3:}$ are symmetric from the geometry, we do not plot in $\mathbf{q} _{3:}$.

From \autoref{fig:experiments_pureflow_4bead_stress}, the major component of the polymer stress $\boldsymbol{\tau}$ is the $x _{1} x _{1}$ entry, and it grows with time, which means the representative particles keep stretching in the $x _{1}$ direction.

The initial state and the final state are quite different; moreover, both representative particles $\mathbf{q}$ and polymer stress $\boldsymbol{\tau}$ continue to change after $t = 3$, it makes the prediction between $t = 3$ and $t = 6$ from the data collected between $t = 0$ and $t = 3$ non-trivial.

\begin{figure}[!ht]
  \centering
  \includegraphics[width = 0.49\textwidth]{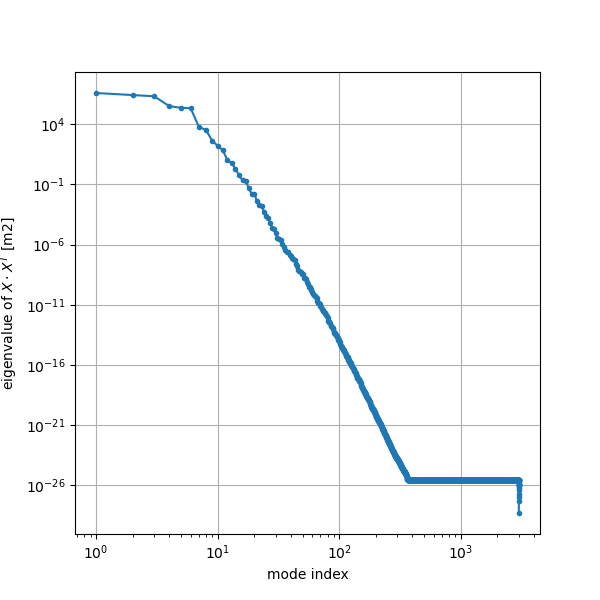}
  \includegraphics[width = 0.49\textwidth]{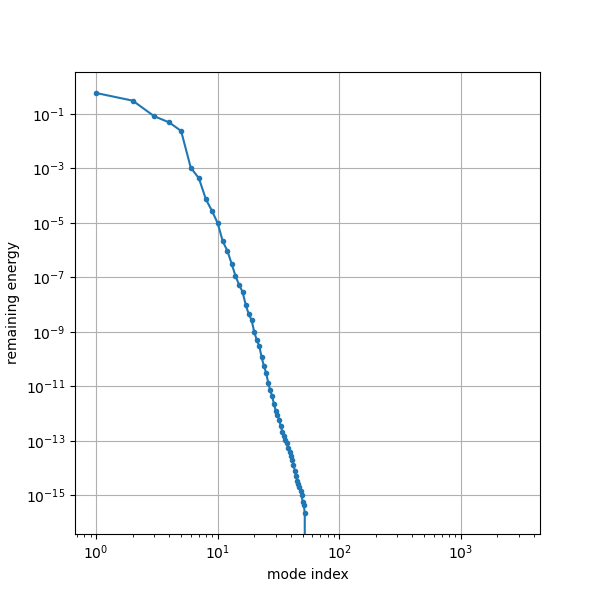}
  \caption{Eigenvalues of $\mathbf{X} \cdot \mathbf{X} \tp$ collected from the reference dynamics. The left part shows the eigenvalues $\lbk \lambda _{i} \rbk$ vs. mode index $i = 1, \cdots, \[ N - 1 \] L$. The right part shows the remaining energy $\lbk 1 - \sum _{j = 1} ^{i} \lambda _{j} / \sum _{k = 1} ^{L} \lambda _{k} \rbk$ vs. mode index $i = 1, \cdots, \[ N - 1 \] L$. Parameters: $4$-bead (homogeneous bonds) polymer molecules with simple shear flow, particle number $P = 1000$, time step size $\delta _{t} = 0.001$, number of snapshots $L = 1000$.}
  \label{fig:experiments_pureflow_4bead_eigen}
\end{figure}

For $4$-bead polymer molecules with $P = 1000$ representative particles and $L = 1000$ snapshots, we plot the decay of eigenvalues of $\mathbf{X} \cdot \mathbf{X} \tp$ (diagonal values of $\hat{\boldsymbol{\Sigma}}$, the $i$-th largest one is denoted as $\lbk \lambda _{i} \rbk$) and the remaining energy (the $i$-th one is denoted as $1 - \sum _{j = 1} ^{i} \lambda _{j} / \sum _{k = 1} ^{L} \lambda _{k}$, the energy fraction that not included in the first $i$ modes) in \autoref{fig:experiments_pureflow_4bead_eigen}.

From the data spanning time $t = 0$ to $t = 3$, a rapid decay is observed between the $6$th and $7$th eigenvalues, followed by an exponential convergence of subsequent eigenvalues (see the left part of \autoref{fig:experiments_pureflow_4bead_eigen}). From the energy perspective, the $7$th and higher eigenvalues collectively contribute only $0.1 \%$ of the total energy (see the right part of \autoref{fig:experiments_pureflow_4bead_eigen}). This suggests that using $6$ POD modes could be a natural choice for constructing a POD-MOR. However, since the nonlinear dynamical system to be predicted covers a longer duration from $t = 0$ to $t = 6$, it is necessary to include a slightly larger number of modes. On the other hand, considering the residual energy, eigenvalues beyond the $50$th have energies that approaching to machine precision. Hence, we do not select more than $50$ POD modes. For this reason, we pick the reduced DoF $R$ from $10$ to $40$ in the following numerical experiments.

We take $R = 10, 15, 20, 25, 30, 40$, and we focus on the computational time and MOR error. Some of the results are listed in \autoref{tab:experiments_shearflow_2Bead_P1k}, \autoref{tab:experiments_shearflow_4Bead_P1k}, \autoref{tab:experiments_shearflow_4Bead_P10k} for $2$-bead $P = 1000$, $4$-bead $P = 1000$, and $4$-bead $P = 10000$, respectively.

\begin{table}[!ht]
  \centering
  \begin{tabular}{|c|c|c|c|c|c|c|}
    \hline
    \multicolumn{7}{|c|}{Full model: $T ^{\textup{ref}} = 385$ (seconds), $F = 3000$.}
    \\
    \hline
    \hline
    $R$ & $10$ & $15$ & $20$ & $25$ & $30$ & $40$
    \\
    \hline
    $T ^{\textup{mor}} / T ^{\textup{ref}}$ & $3.77\%$ & $7.31\%$ & $12.2\%$ & $18.4\%$ & $26.0\%$ & $44.9\%$
    \\
    \hline
    $\textup{err} _{\textup{MOR}} \( t = 6 \)$ & $10.5\%$ & $9.47\%$ & $8.76\%$ & $8.41\%$ & $8.23\%$ & $8.08\%$
    \\
    \hline
  \end{tabular}
  \caption{The relation among the reduced DoF $R$, the ratio of computational time $T ^{\textup{mor}} / T ^{\textup{ref}}$, and the relative $L ^{2}$ MOR error $\textup{err} _{\textup{MOR}} \( t = 6 \)$, for different reduced DoF $R$ in the POD-MOR dynamics. Parameters: $2$-bead polymer molecules with simple shear flow, particle number $P = 1000$, time step size $\delta _{t} = 0.001$, number of snapshots $L = 3000$.}
  \label{tab:experiments_shearflow_2Bead_P1k}
\end{table}

\begin{table}[!ht]
  \centering
  \begin{tabular}{|c|c|c|c|c|c|c|}
    \hline
    \multicolumn{7}{|c|}{Full model: $T ^{\textup{ref}} = 704$ (seconds), $F = 9000$.}
    \\
    \hline
    \hline
    $R$ & $10$ & $15$ & $20$ & $25$ & $30$ & $40$
    \\
    \hline
    $T ^{\textup{mor}} / T ^{\textup{ref}}$ & $2.30\%$ & $4.24\%$ & $6.89\%$ & $10.3\%$ & $14.4\%$ & $24.8\%$
    \\
    \hline
    $\textup{err} _{\textup{MOR}} \( t = 6 \)$ & $12.5\%$ & $6.67\%$ & $5.67\%$ & $5.00\%$ & $4.42\%$ & $3.72\%$
    \\
    \hline
  \end{tabular}
  \caption{The relation among the reduced DoF $R$, the ratio of computational time $T ^{\textup{mor}} / T ^{\textup{ref}}$, and the relative $L ^{2}$ MOR error $\textup{err} _{\textup{MOR}} \( t = 6 \)$, for different reduced DoF $R$ in the POD-MOR dynamics. Parameters: $4$-bead (homogeneous bonds) polymer molecules with simple shear flow, particle number $P = 1000$, time step size $\delta _{t} = 0.001$, number of snapshots $L = 1000$.}
  \label{tab:experiments_shearflow_4Bead_P1k}
\end{table}

\begin{table}[!ht]
  \centering
  \begin{tabular}{|c|c|c|c|c|c|c|}
    \hline
    \multicolumn{7}{|c|}{Full model: $T ^{\textup{ref}} = 70104$ (seconds), $F = 90000$.}
    \\
    \hline
    \hline
    $R$ & $10$ & $15$ & $20$ & $25$ & $30$ & $40$
    \\
    \hline
    $T ^{\textup{mor}} / T ^{\textup{ref}}$ & $2.31\%$ & $4.30\%$ & $6.90\%$ & $10.5\%$ & $14.8\%$ & $25.4\%$
    \\
    \hline
    $\textup{err} _{\textup{MOR}} \( t = 6 \)$ & $13.2\%$ & $6.60\%$ & $5.99\%$ & $5.59\%$ & $5.41\%$ & $5.18\%$
    \\
    \hline
  \end{tabular}
  \caption{The relation among the reduced DoF $R$, the ratio of computational time $T ^{\textup{mor}} / T ^{\textup{ref}}$, and the relative $L ^{2}$ MOR error $\textup{err} _{\textup{MOR}} \( t = 6 \)$, for different reduced DoF $R$ in the POD-MOR dynamics. Parameters: $4$-bead (homogeneous bonds) polymer molecules with simple shear flow, particle number $P = 10000$, time step size $\delta _{t} = 0.001$, number of snapshots $L = 10000$.}
  \label{tab:experiments_shearflow_4Bead_P10k}
\end{table}

The numerical experiments in \autoref{sec:accuracy} show the numerical error of the reference solution of the $4$-bead polymer molecule system with $P = 1000$ in simple shear flow is about $5\% \sim 10\%$ relatively, which is regarded as the benchmark of the POD-MOR error, see the discussion in \autoref{sec:accuracy}. That means the above POD-MOR error shown in \autoref{tab:experiments_shearflow_2Bead_P1k}, \autoref{tab:experiments_shearflow_4Bead_P1k}, \autoref{tab:experiments_shearflow_4Bead_P10k} as well as other cases is at the same level as the errors of the reference solution, which implies that the POD-MOR provides a good approximation with less computational cost. These results show that the POD-MOR is more efficient than the nonlinear reference dynamics while maintaining sufficient accuracy.

\begin{figure}[!ht]
  \centering
  \includegraphics[width = 0.5\textwidth]{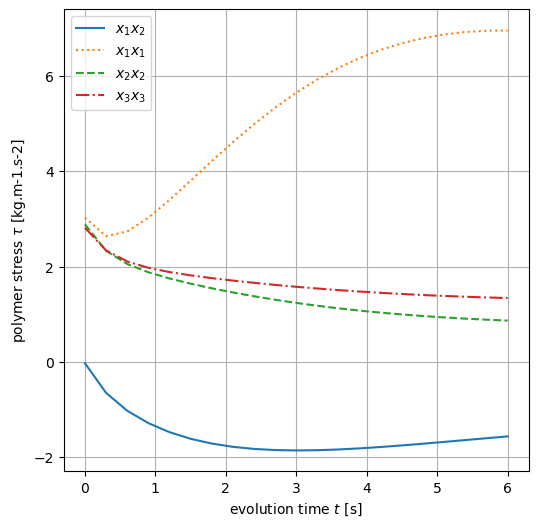}
  \caption{Polymer stress $\boldsymbol{\tau}$ vs. evolution time $t$ in the reference dynamics. Parameters: $4$-bead (inhomogeneous bonds) polymer molecules with simple shear flow, particle number $P = 1000$, time step size $\delta _{t} = 0.001$.}
  \label{fig:experiments_pureflow_4beadH_stress}
\end{figure}

\begin{figure}[!ht]
  \centering
  \includegraphics[width = 0.24\textwidth]{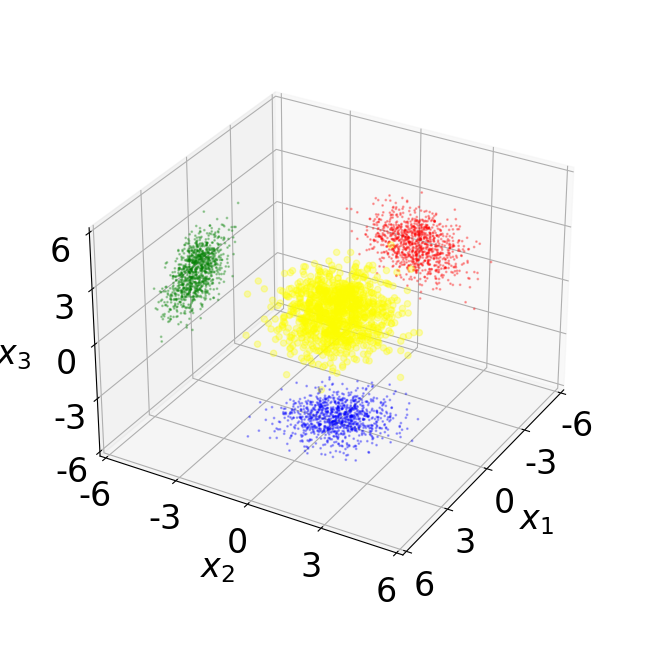}
  \includegraphics[width = 0.24\textwidth]{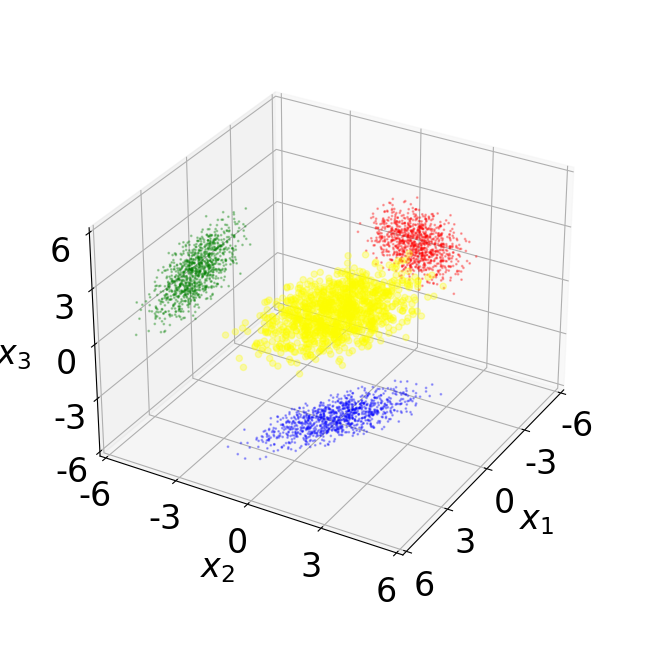}
  \includegraphics[width = 0.24\textwidth]{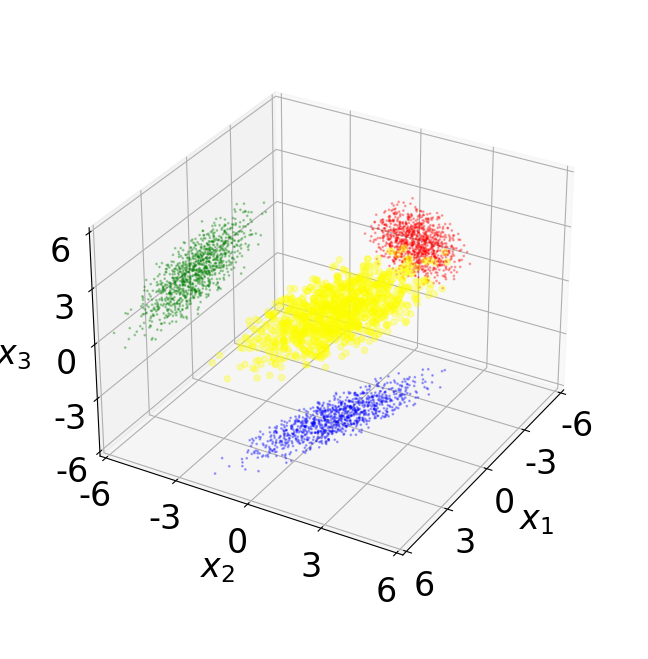}
  \includegraphics[width = 0.24\textwidth]{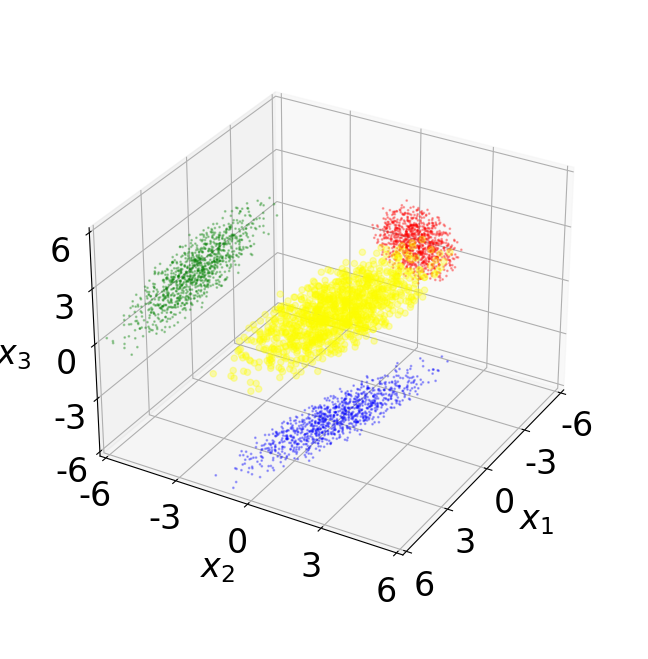}
  \\
  \includegraphics[width = 0.24\textwidth]{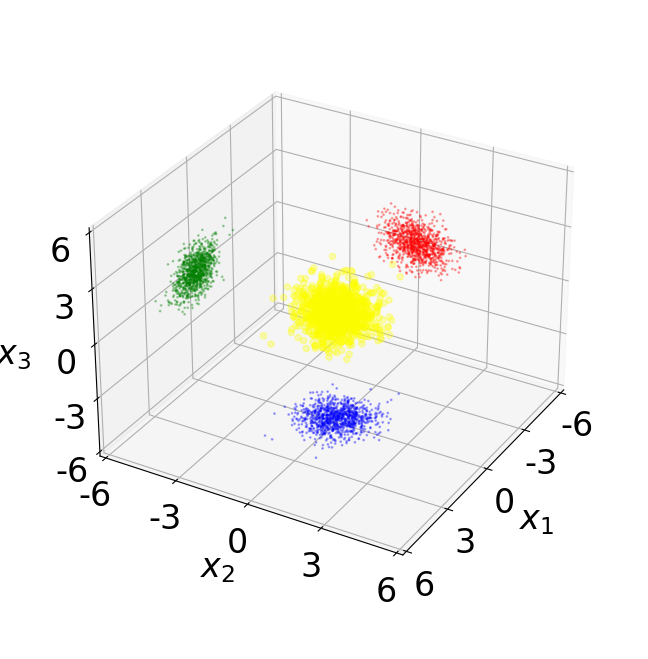}
  \includegraphics[width = 0.24\textwidth]{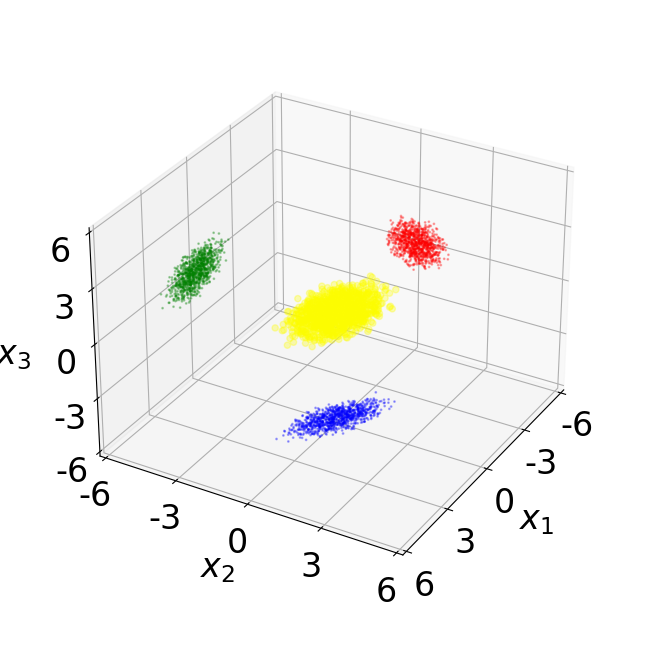}
  \includegraphics[width = 0.24\textwidth]{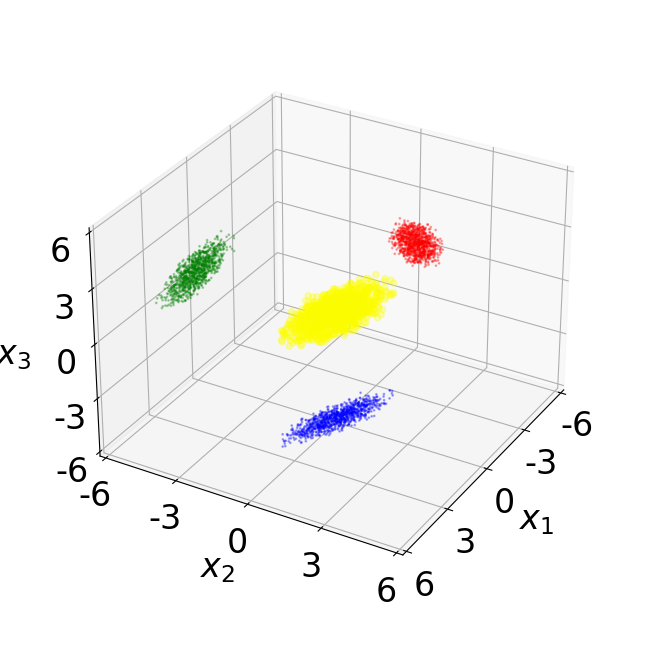}
  \includegraphics[width = 0.24\textwidth]{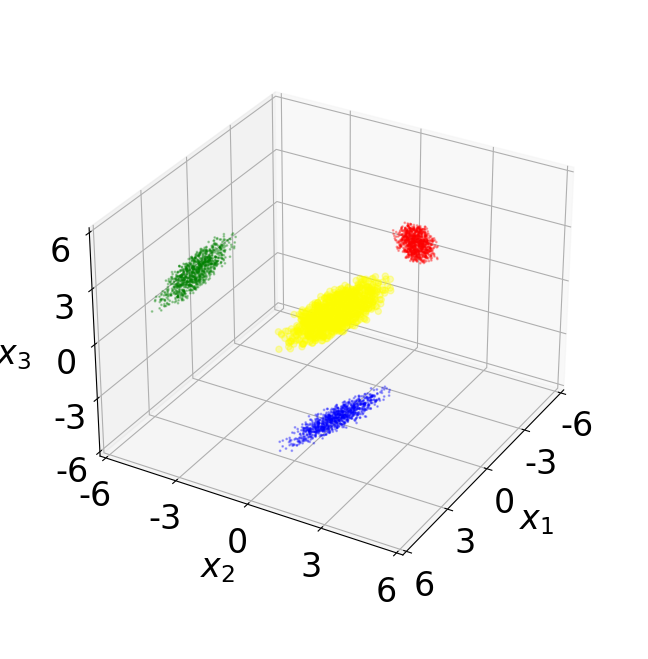}
  \\
  \includegraphics[width = 0.24\textwidth]{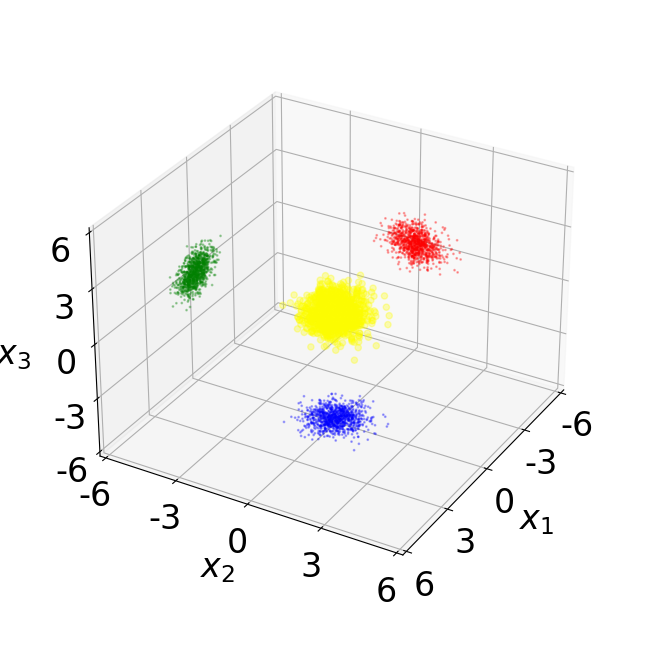}
  \includegraphics[width = 0.24\textwidth]{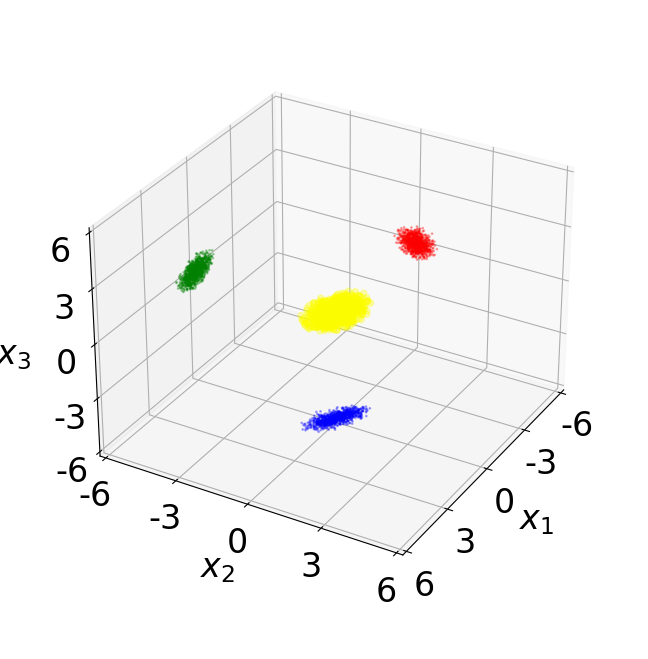}
  \includegraphics[width = 0.24\textwidth]{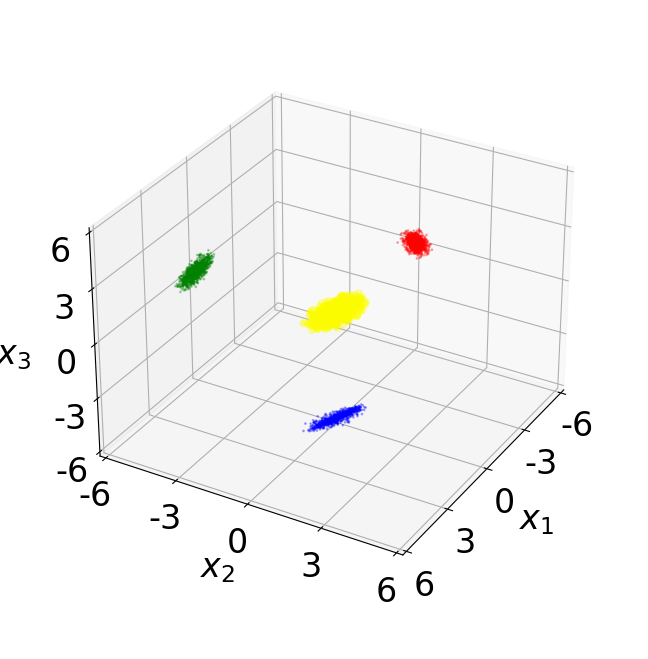}
  \includegraphics[width = 0.24\textwidth]{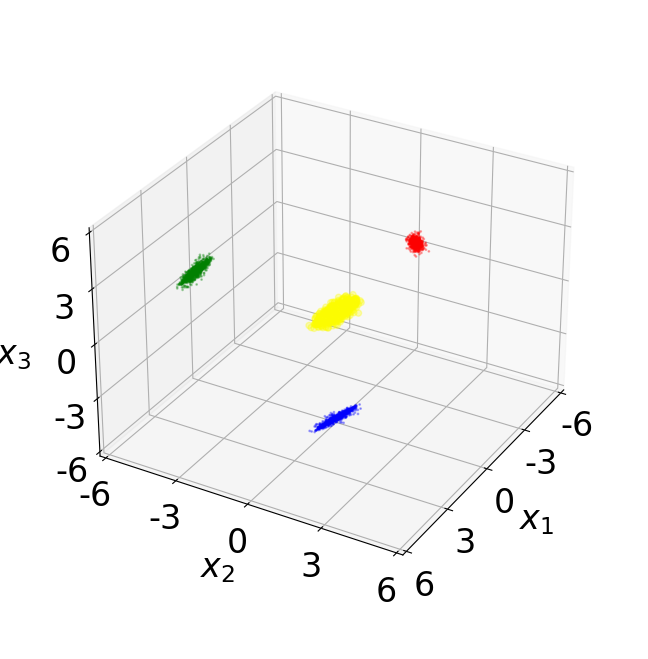}
  \caption{The bond of representative particles $\mathbf{q} _{1:}$, $\mathbf{q} _{2:}$, and $\mathbf{q} _{3:}$ in the reference dynamics. From top to bottom, the plots correspond to the first bond $\mathbf{q} _{1:}$, the second bond $\mathbf{q} _{2:}$, and the third bond $\mathbf{q} _{3:}$. From left to right, the plots correspond to evolution time $t = 0, 1.5, 3, 6 \[ \textup{s} \]$. The bonds are illustrated in 3D space (yellow) and 2D projections on $x _{1} x _{2}$ (blue), $x _{2} x _{3}$ (red), $x _{3} x _{1}$ (green) planes. Parameters: $4$-bead (inhomogeneous bonds) polymer molecules with simple shear flow, particle number $P = 1000$, time step size $\delta _{t} = 0.001$.}
  \label{fig:experiments_pureflow_4beadH_particle}
\end{figure}

For the cases with inhomogeneous bonds, the setting is the same as the homogeneous case.
Again, we plot the polymer stress in \autoref{fig:experiments_pureflow_4beadH_stress} and evolution of the representative particles in \autoref{fig:experiments_pureflow_4beadH_particle} for the system of $4$-bead polymer molecules with $P = 1000$ representative particles.


Since all three bonds have different Hookean constants, the evolutions behave differently on them (see \autoref{fig:experiments_pureflow_4beadH_particle}). More specifically, the bond ($\mathbf{q} _{1:}$) with the smallest Hookean constant $H _{1} = 1$ shows a more spread-out distribution, while the bond ($\mathbf{q} _{3:}$) with the largest Hookean constant $H _{3} = 3$ shows a more concentrated distribution.

The overall Hookean constants are larger compared with the homogeneous case, which means the relaxation time of polymer molecules is shorter, i.e., the evolution of the polymer stress $\boldsymbol{\tau}$ is faster (see \autoref{fig:experiments_pureflow_4beadH_stress}). 

We take $R = 10, 15, 20, 25, 30, 40$, and we list the computational time and MOR error in \autoref{tab:experiments_shearflow_4BeadH_P1k} is for $4$-bead $P = 1000$.
The relative POD-MOR error is higher than the homogeneous bond cases, but it is still in the regime of the benchmark, i.e., $5\% \sim 10\%$, discussed in \autoref{sec:accuracy}, which means the POD-MOR is still efficient than the nonlinear reference dynamics while maintaining sufficient accuracy.

\begin{table}[!ht]
  \centering
  \begin{tabular}{|c|c|c|c|c|c|c|}
    \hline
    \multicolumn{7}{|c|}{Full model: $T ^{\textup{ref}} = 673$ (seconds), $F = 9000$.}
    \\
    \hline
    \hline
    $R$ & $10$ & $15$ & $20$ & $25$ & $30$ & $40$
    \\
    \hline
    $T ^{\textup{mor}} / T ^{\textup{ref}}$ & $2.39\%$ & $4.43\%$ & $7.20\%$ & $10.8\%$ & $15.1\%$ & $25.8\%$
    \\
    \hline
    $\textup{err} _{\textup{MOR}} \( t = 6 \)$ & $18.3\%$ & $11.1\%$ & $9.55\%$ & $8.36\%$ & $7.93\%$ & $7.02\%$
    \\
    \hline
  \end{tabular}
  \caption{The relation among the reduced DoF $R$, the ratio of computational time $T ^{\textup{mor}} / T ^{\textup{ref}}$, and the relative $L ^{2}$ MOR error $\textup{err} _{\textup{MOR}} \( t = 6 \)$, for different reduced DoF $R$ in the POD-MOR dynamics. Parameters: $4$-bead (inhomogeneous bonds) polymer molecules with simple shear flow, particle number $P = 1000$, time step size $\delta _{t} = 0.001$, number of snapshots $L = 1000$.}
  \label{tab:experiments_shearflow_4BeadH_P1k}
\end{table}

\begin{figure}[!ht]
  \centering
  \includegraphics[width = 0.49\textwidth]{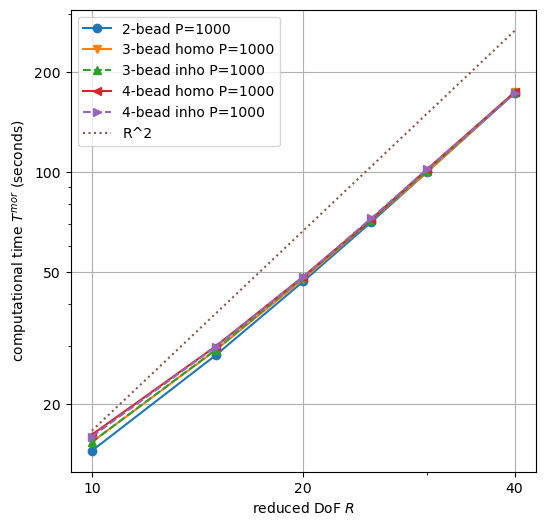}
  \includegraphics[width = 0.49\textwidth]{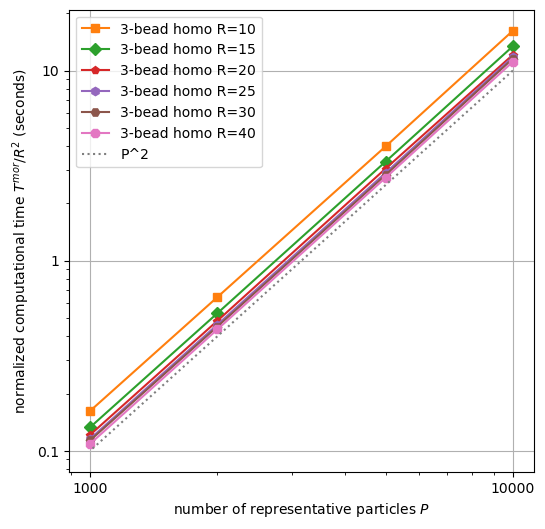}
  \caption{The computation time in the POD-MOR dynamics. The left part shows computational time $T ^{\textup{mor}}$ (seconds) vs. reduced DoF $R$. Parameters: simple shear flow, particle number $P = 1000$, time step size $\delta _{t} = 0.001$. The right part shows normalized computational time $T ^{\textup{mor}} / R ^{2}$ (seconds) vs. number of representative particles $P$. Parameters: $3$-bead (homogeneous bonds) polymer molecules with simple shear flow, time step size $\delta _{t} = 0.001$.}
  \label{fig:experiments_quadratic}
\end{figure}

From the above numerical experiments, we observe that the computational time is a quadratic function of reduced DoF $T ^{\textup{mor}} \sim R ^{2}$ when we fix $P = 1000$, and it is not sensitive to the complexity of the polymer structure, see the left part of \autoref{fig:experiments_quadratic}. Actually, the computational time is a quadratic function of both reduced DoF and number of representative particles $T ^{\textup{mor}} \sim P ^{2} R ^{2}$, as shown in the right part of \autoref{fig:experiments_quadratic}. This is mainly because computational cost on the evolution reduced system is neglectable compared with the cost on the POD mappings \autoref{eq:model_pq}.

\begin{figure}[!ht]
  \centering
  \includegraphics[width = 0.49\textwidth]{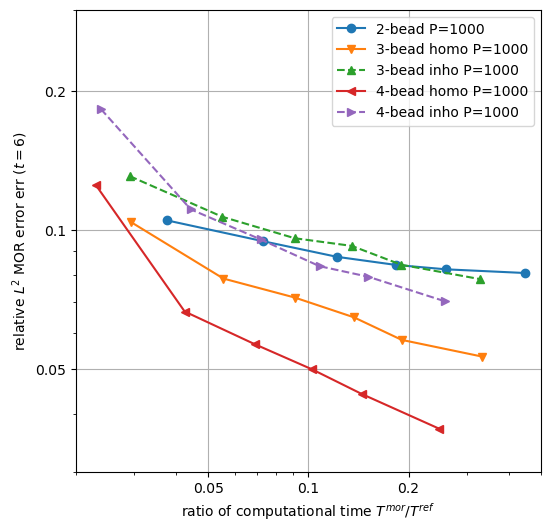}
  \includegraphics[width = 0.49\textwidth]{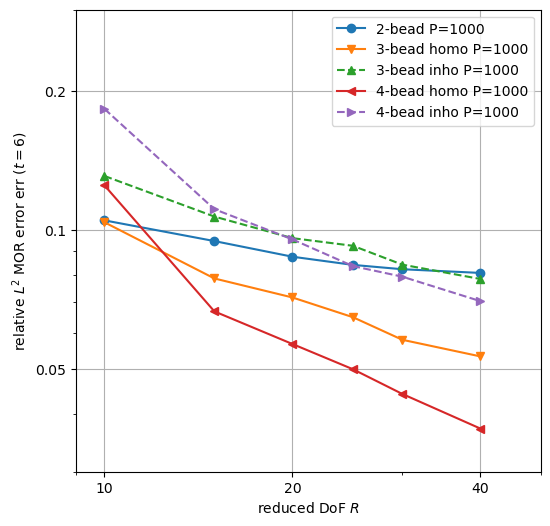}
  \caption{The MOR error in the POD-MOR dynamics. The left part shows relative $L ^{2}$ MOR error $\textup{err} _{\textup{MOR}} \( t = 6 \)$ vs. ratio of computational time $T ^{\textup{mor}} / T ^{\textup{ref}}$. Parameters: simple shear flow, particle number $P = 1000$, time step size $\delta _{t} = 0.001$. The right part shows relative $L ^{2}$ MOR error $\textup{err} _{\textup{MOR}} \( t = 6 \)$ vs. reduced DoF $R$. Parameters: the same as the left part.}
  \label{fig:experiments_Err_TT_P1k}
\end{figure}

One of the most important observations is that when the number of representative particles $P$ is fixed, the saving from MOR is greater for higher molecular complexity, see \autoref{fig:experiments_Err_TT_P1k}. 
It shows that the POD-MOR method may be useful and helpful in solving problems with high computational complexity. In the left part of \autoref{fig:experiments_Err_TT_P1k}, the relative computational time $T ^{\textup{mor}} / T ^{\textup{ref}} = 1$ means the reduced model costs the same computational time as the full model, and the POD-MOR saves time if the relative computational time $T ^{\textup{mor}} / T ^{\textup{ref}} < 1$.

\begin{figure}[!ht]
  \centering
  \includegraphics[width = 0.49\textwidth]{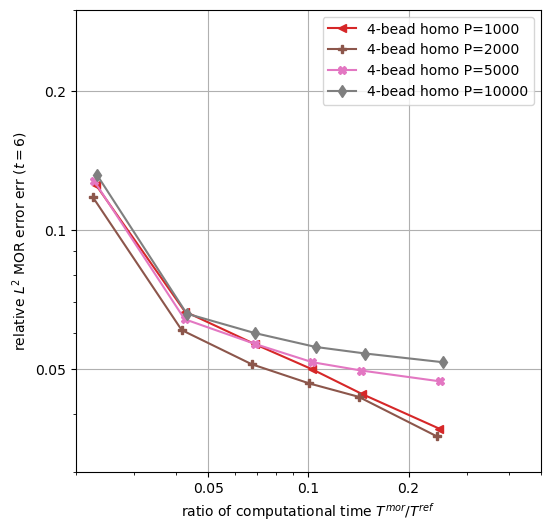}
  \includegraphics[width = 0.49\textwidth]{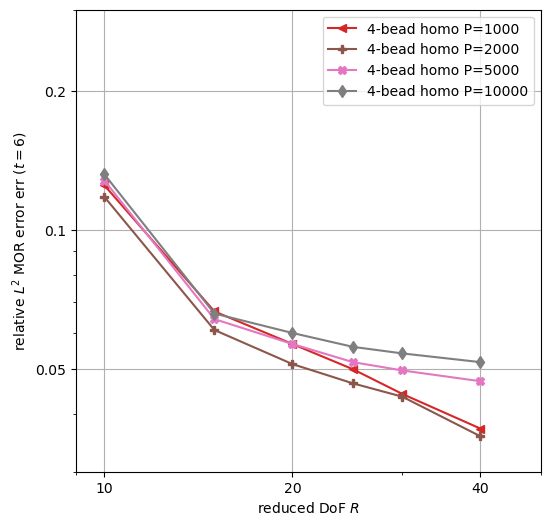}
  \caption{The MOR error in the POD-MOR dynamics. The left part shows relative $L ^{2}$ MOR error $\textup{err} _{\textup{MOR}} \( t = 6 \)$ vs. ratio of computational time $T ^{\textup{mor}} / T ^{\textup{ref}}$. Parameters: $4$-bead (homogeneous bonds) polymer molecules with simple shear flow, time step size $\delta _{t} = 0.001$. The right part shows relative $L ^{2}$ MOR error $\textup{err} _{\textup{MOR}} \( t = 6 \)$ vs. reduced DoF $R$. Parameters: the same as the left part.}
  \label{fig:experiments_Err_TT_3BondHomo}
\end{figure}

Increasing the number of representative particles $P$ results in a larger POD-MOR error by a few percent, see \autoref{fig:experiments_Err_TT_3BondHomo}. Nevertheless, the difference between $P = 1000$ and $P = 10000$ is not quite large (compared with the gap between $3$-bead and $4$-bead in \autoref{fig:experiments_Err_TT_P1k}), especially when $R$ is small suggesting that POD-MOR is still effective for huge systems with large amount of representative particles.

Again, among almost all the tests in both \autoref{fig:experiments_Err_TT_P1k} and \autoref{fig:experiments_Err_TT_3BondHomo}, the relative POD-MOR error is at the same level as the benchmark $5\% \sim 10\%$ discussed in \autoref{sec:accuracy}, which means the POD-MOR accelerate the computation without generating substantial additional error.

The inhomogeneity of the bond coefficients also results in a larger relative error by a few percent. This is reasonable since we use the shared POD space for all bonds regardless of their intrinsic bond coefficients and their different responses on the shear flows. Nevertheless, the POD-MOR behaves slightly better on $4$-bead than on $3$-bead when the bond coefficients are inhomogeneous.

\section{Conclusions}
\label{sec:conclusions}

In this paper, we solve the microscopic Fokker-Planck equation using VDS and further apply POD-MOR to reduce computational time.

The VDS is extended to 3D and multi-bead setups, which is much closer to the practical applications. Numerical experiments validate the feasibility of POD-MOR in these cases since the relative POD-MOR error is at the same level as the relative errors of the reference dynamics, which is $5\% \sim 10\%$, and reduces the computational time.

From our numerical experiments, the no flow and the simple shear flows are implemented. Starting from the thermal equilibrium, we need few POD modes to achieve an accurate reduced model in the no flow process. For example, under simple shear flows, the reduced-order model introduces about $6\%$ relative error in predicting the dynamics while requiring only a fraction (about $6\%$) of the original computational time for $4$-bead chain polymers, and the degrees of freedom can be reduced significantly to about $0.1\%$ of the original model, which means the low-dimensional structure is found by POD.
We show that POD-MOR behaves more efficiently on more complicated systems, e.g., $4$-bead polymer molecules compared with $3$-bead polymer molecules, even if we choose a larger number of representative particles for $4$-bead system. The advantage of POD-MOR for higher molecular complexity still exists when inhomogeneous potential coefficients are introduced on polymer bonds.

This establishes a practical pathway for multiscale and complex fluid simulations.

\section*{Acknowledgments}

We would like to acknowledge the beneficial discussion with Faisal Habib. This work is supported in part by the Natural Sciences and Engineering Research Council of Canada and the Fields Institute MADS Lab.

\section*{Code Availability}

The custom code used in this work is available upon request.

\bibliographystyle{siamplain}
\bibliography{notes_DVPPOD}

@article{fang0000proper1,
  title={Proper Orthogonal Decomposition-Based Model-Order Reduction for Smoothed Particle Hydrodynamics Simulation},
  author={Fang, Lidong and Song, Zilong and Fraser, Kirk and Habib, Faisal and Drummond, Christopher and Huang, Huaxiong},
  journal={arXiv preprint arXiv:2507.19825},
  year={2025}
}

@article{fang0000proper2,
  title={Proper Orthogonal Decomposition-based Model-Order Reduction for Smoothed Particle Hydrodynamics Simulation — Mass-Spring-Damper System},
  author={Fang, Lidong and Song, Zilong and Fraser, Kirk and Huang, Huaxiong},
  journal={arXiv preprint arXiv:2508.00335},
  year={2025}
}

@article{cao2022machine,
  title={Machine Learning and Reduced Order Computation of a Friction Stir Welding Model},
  author={Cao, Xiulei and Fraser, Kirk and Song, Zilong and Drummond, Chris and Huang, Huaxiong},
  journal={Journal of Computational Physics},
  volume={454},
  pages={110863},
  year={2022},
  publisher={Elsevier}
}

@article{lassila2014model,
  title={Model Order Reduction in Fluid Dynamics: Challenges and Perspectives},
  author={Lassila, Toni and Manzoni, Andrea and Quarteroni, Alfio and Rozza, Gianluigi},
  journal={Reduced Order Methods for Modeling and Computational Reduction},
  pages={235--273},
  year={2014},
  publisher={Springer}
}

@article{lucia2004reduced,
  title={Reduced-Order Modeling: New Approaches for Computational Physics},
  author={Lucia, David J and Beran, Philip S and Silva, Walter A},
  journal={Progress in Aerospace Sciences},
  volume={40},
  number={1-2},
  pages={51--117},
  year={2004},
  publisher={Elsevier}
}

@article{peherstorfer2015dynamic,
  title={Dynamic Data-Driven Reduced-Order Models},
  author={Peherstorfer, Benjamin and Willcox, Karen},
  journal={Computer Methods in Applied Mechanics and Engineering},
  volume={291},
  pages={21--41},
  year={2015},
  publisher={Elsevier}
}

@article{taira2017modal,
  title={Modal Analysis of Fluid Flows: An Overview},
  author={Taira, Kunihiko and Brunton, Steven L and Dawson, Scott TM and Rowley, Clarence W and Colonius, Tim and McKeon, Beverley J and Schmidt, Oliver T and Gordeyev, Stanislav and Theofilis, Vassilios and Ukeiley, Lawrence S},
  journal={AIAA Journal},
  volume={55},
  number={12},
  pages={4013--4041},
  year={2017},
  publisher={American Institute of Aeronautics and Astronautics}
}

@article{benner2015survey,
  title={A Survey of Projection-Based Model Reduction Methods for Parametric Dynamical Systems},
  author={Benner, Peter and Gugercin, Serkan and Willcox, Karen},
  journal={SIAM Review},
  volume={57},
  number={4},
  pages={483--531},
  year={2015},
  publisher={SIAM}
}

@article{copeland2022reduced,
  title={Reduced Order Models for Lagrangian Hydrodynamics},
  author={Copeland, Dylan Matthew and Cheung, Siu Wun and Huynh, Kevin and Choi, Youngsoo},
  journal={Computer Methods in Applied Mechanics and Engineering},
  volume={388},
  pages={114259},
  year={2022},
  publisher={Elsevier}
}

@inproceedings{magargal2022lagrangian,
  title={Lagrangian-to-Eulerian Mapping for Modal Decomposition of Multiphase Flows via Smoothed-Particle Hydrodynamics},
  author={Magargal, Liam K and Rodriguez, Steven N and Jaworski, Justin and Iliopoulos, Athanasios P and Michopoulos, John G},
  booktitle={AIAA AVIATION 2022 Forum},
  pages={4169},
  year={2022}
}

@article{mojgani2017lagrangian,
  title={Lagrangian Basis Method for Dimensionality Reduction of Convection Dominated Nonlinear Flows},
  author={Mojgani, Rambod and Balajewicz, Maciej},
  journal={arXiv preprint arXiv:1701.04343},
  year={2017}
}

@article{chatterjee2000introduction,
  title={An Introduction to the Proper Orthogonal Decomposition},
  author={Chatterjee, Anindya},
  journal={Current Science},
  pages={808--817},
  year={2000},
  publisher={JSTOR}
}

@article{berkooz1993proper,
  title={The Proper Orthogonal Decomposition in the Analysis of Turbulent Flows},
  author={Berkooz, Gal and Holmes, Philip and Lumley, John L},
  journal={Annual Review of Fluid Mechanics},
  volume={25},
  number={1},
  pages={539--575},
  year={1993},
  publisher={Annual Reviews}
}

@article{kerschen2005method,
  title={The Method of Proper Orthogonal Decomposition for Dynamical Characterization and Order Reduction of Mechanical Systems: An Overview},
  author={Kerschen, Gaetan and Golinval, Jean-claude and Vakakis, Alexander F and Bergman, Lawrence A},
  journal={Nonlinear Dynamics},
  volume={41},
  pages={147--169},
  year={2005},
  publisher={Springer}
}

@article{rapun2010reduced,
  title={Reduced Order Models Based on Local POD Plus Galerkin Projection},
  author={Rap{\'u}n, Mar{\'\i}a-Luisa and Vega, Jos{\'e} M},
  journal={Journal of Computational Physics},
  volume={229},
  number={8},
  pages={3046--3063},
  year={2010},
  publisher={Elsevier}
}

@article{carlberg2017galerkin,
  title={Galerkin v. Least-Squares Petrov--Galerkin Projection in Nonlinear Model Reduction},
  author={Carlberg, Kevin and Barone, Matthew and Antil, Harbir},
  journal={Journal of Computational Physics},
  volume={330},
  pages={693--734},
  year={2017},
  publisher={Elsevier}
}

@article{choi2019space,
  title={Space--Time Least-Squares Petrov--Galerkin Projection for Nonlinear Model Reduction},
  author={Choi, Youngsoo and Carlberg, Kevin},
  journal={SIAM Journal on Scientific Computing},
  volume={41},
  number={1},
  pages={A26--A58},
  year={2019},
  publisher={SIAM}
}

@article{carlberg2011efficient,
  title={Efficient Non-Linear Model Reduction via a Least-Squares Petrov--Galerkin Projection and Compressive Tensor Approximations},
  author={Carlberg, Kevin and Bou-Mosleh, Charbel and Farhat, Charbel},
  journal={International Journal for Numerical Methods in Engineering},
  volume={86},
  number={2},
  pages={155--181},
  year={2011},
  publisher={Wiley Online Library}
}

@book{bird1987dynamics,
  title={Dynamics of polymeric liquids, volume 2: Kinetic theory},
  author={Bird, Robert Byron and Curtiss, Charles F and Armstrong, Robert C and Hassager, Ole},
  year={1987},
  publisher={Wiley}
}

@article{laso1993calculation,
  title={Calculation of viscoelastic flow using molecular models: the CONNFFESSIT approach},
  author={Laso, Manuel and {\"O}ttinger, Hans Christian},
  journal={Journal of Non-Newtonian Fluid Mechanics},
  volume={47},
  pages={1--20},
  year={1993},
  publisher={Elsevier}
}

@article{le2012micro,
  title={Micro-macro models for viscoelastic fluids: modelling, mathematics and numerics},
  author={Le Bris, Claude and Lelievre, Tony},
  journal={Science China Mathematics},
  volume={55},
  pages={353--384},
  year={2012},
  publisher={Springer}
}

@article{fang2022deepn,
  title={DeePN$2$: A Deep Learning-Based Non-Newtonian Hydrodynamic Model},
  author={Fang, Lidong and Ge, Pei and Zhang, Lei and others},
  journal={Journal of Machine Learning},
  volume={1},
  number={1},
  pages={114--140},
  year={2022},
  publisher={Global Science Press}
}

@book{ottinger2012stochastic,
  title={Stochastic processes in polymeric fluids: tools and examples for developing simulation algorithms},
  author={{\"O}ttinger, Hans C},
  year={2012},
  publisher={Springer Science \& Business Media}
}

@article{lin2007micro,
  title={On a micro-macro model for polymeric fluids near equilibrium},
  author={Lin, Fang-Hua and Liu, Chun and Zhang, Ping},
  journal={Communications on Pure and Applied Mathematics: A Journal Issued by the Courant Institute of Mathematical Sciences},
  volume={60},
  number={6},
  pages={838--866},
  year={2007},
  publisher={Wiley Online Library}
}

@article{wang2021two,
  title={A two species micro--macro model of wormlike micellar solutions and its maximum entropy closure approximations: An energetic variational approach},
  author={Wang, Yiwei and Zhang, Teng-Fei and Liu, Chun},
  journal={Journal of Non-Newtonian Fluid Mechanics},
  volume={293},
  pages={104559},
  year={2021},
  publisher={Elsevier}
}

@article{bao2024micro,
  title={Micro-macro modeling of polymeric fluids and shear-induced microscopic behaviors with bond-breaking},
  author={Bao, Xuelian and Huang, Huaxiong and Song, Zilong and Xu, Shixin},
  journal={Physical Review Fluids},
  volume={9},
  number={10},
  pages={103301},
  year={2024},
  publisher={APS}
}

@article{bao2025deterministic,
  title={A deterministic--particle--based scheme for micro-macro viscoelastic flows},
  author={Bao, Xuelian and Liu, Chun and Wang, Yiwei},
  journal={Journal of Computational Physics},
  volume={522},
  pages={113589},
  year={2025},
  publisher={Elsevier}
}

@article{bao2025micro,
  title={Micro-Macro Modeling of Polymeric Fluids with Multi-Bead Polymer Chain},
  author={Bao, Xuelian and Fang, Lidong and Huang, Huaxiong and Song, Zilong and Xu, Shixin},
  journal={arXiv preprint arXiv:2506.08377},
  year={2025}
}

@article{pearson1901liii,
  title={LIII. On lines and planes of closest fit to systems of points in space},
  author={Pearson, Karl},
  journal={The London, Edinburgh, and Dublin philosophical magazine and journal of science},
  volume={2},
  number={11},
  pages={559--572},
  year={1901},
  publisher={Taylor \& Francis}
}

@article{hotelling1933analysis,
  title={Analysis of a complex of statistical variables into principal components.},
  author={Hotelling, Harold},
  journal={Journal of educational psychology},
  volume={24},
  number={6},
  pages={417},
  year={1933},
  publisher={Warwick \& York}
}

@article{loeve1945calcul,
  title={Calcul des probabilit{\'e}s--sur la covariance d’une fonction al{\'e}atoire (Cal-culating probabilities--on the covariance of a random function). Note by M. Michel Lo{\'e}ve, present by M. Henri Villat},
  author={Lo{\`e}ve, Michel},
  journal={Comptes rendus de l’Acad{\'e}mie des Sciences (Comptes rendus, or Proceedings of the Academy of sciences)},
  volume={220},
  pages={295--296},
  year={1945}
}

@article{karhunen1946spektraltheorie,
  title={Zur spektraltheorie stochastischer prozesse},
  author={Karhunen, Kari},
  journal={Ann. Acad. Sci. Fennicae, AI},
  volume={34},
  year={1946}
}

@article{sirovich1987turbulence1,
  title={Turbulence and the dynamics of coherent structures. I. Coherent structures},
  author={Sirovich, Lawrence},
  journal={Quarterly of applied mathematics},
  volume={45},
  number={3},
  pages={561--571},
  year={1987}
}

@article{sirovich1987turbulence2,
  title={Turbulence and the dynamics of coherent structures. II. Symmetries and transformations},
  author={Sirovich, Lawrence},
  journal={Quarterly of Applied mathematics},
  volume={45},
  number={3},
  pages={573--582},
  year={1987}
}

@article{sirovich1987turbulence3,
  title={Turbulence and the dynamics of coherent structures. III. Dynamics and scaling},
  author={Sirovich, Lawrence},
  journal={Quarterly of Applied mathematics},
  volume={45},
  number={3},
  pages={583--590},
  year={1987}
}

@book{holmes1996turbulence,
  title={Turbulence, Coherent Structures, Dynamical Systems and Symmetry},
  author={Holmes, Philip and Lumley, John L and Berkooz, Gal},
  year={1996},
  publisher={Cambridge University Press}
}

@article{lall2003structure,
  title={Structure-preserving model reduction for mechanical systems},
  author={Lall, Sanjay and Krysl, Petr and Marsden, Jerrold E},
  journal={Physica D: Nonlinear Phenomena},
  volume={184},
  number={1-4},
  pages={304--318},
  year={2003},
  publisher={Elsevier}
}

@article{rathinam2003new,
  title={A new look at proper orthogonal decomposition},
  author={Rathinam, Muruhan and Petzold, Linda R},
  journal={SIAM Journal on Numerical Analysis},
  volume={41},
  number={5},
  pages={1893--1925},
  year={2003},
  publisher={SIAM}
}

@article{rowley2004model,
  title={Model reduction for compressible flows using POD and Galerkin projection},
  author={Rowley, Clarence W and Colonius, Tim and Murray, Richard M},
  journal={Physica D: Nonlinear Phenomena},
  volume={189},
  number={1-2},
  pages={115--129},
  year={2004},
  publisher={Elsevier}
}

@article{rowley2005model,
  title={Model reduction for fluids, using balanced proper orthogonal decomposition},
  author={Rowley, Clarence W},
  journal={International Journal of Bifurcation and Chaos},
  volume={15},
  number={03},
  pages={997--1013},
  year={2005},
  publisher={World Scientific}
}

\appendix

\section{Accuracy Test of the Reference Model}
\label{sec:accuracy}

In this section, we provide the convergence test of the reference model \autoref{eq:model_DymRepPar} to verify the accuracy. We study the convergence on different numerical schemes, different time step sizes, and different number of representative particles.

Here we take the 3D $4$-bead polymer molecule system in the simple shear flow and homogeneous Hookean potential as an example. The setup of the model, including the thermal equilibrium initial condition and the friction coefficient $\zeta = 4$, is the same as in \autoref{sec:experiments}. The total simulation starts at $t = 0$ and ends at time $t = 6$.

We first implement the reference model \autoref{eq:model_DymRepPar} using the first order explicit scheme, i.e., given current simulation time $t$, and define $t ^{+} := t + \delta _{t}$,
\begin{align*}
  &
  \quad
  \frac{1}{\delta _{t}} \[ \bar{\mathbf{q}} _{k, I} \( t ^{+} \) - \bar{\mathbf{q}} _{k, I} \( t \) \]
  +
  \mathbf{u} \cdot \nabla _{\mathbf{x}} \bar{\mathbf{q}} _{k, I} \( t \)
  -
  \bar{\mathbf{q}} _{k, I} \( t \) \cdot \nabla _{\mathbf{x}} \mathbf{u}
  \nonumber
  \\
  &
  =
  -
  \frac{\kB T}{\zeta}
  \sum _{j = 1} ^{N - 1}
  A _{kj}
  \[ \frac{\sum _{K} \nabla _{\mathbf{q} _{j}} K _{h} \( \bar{\mathbf{q}} _{I} \( t \) - \bar{\mathbf{q}} _{K} \( t \) \)}{\sum _{J} K _{h} \( \bar{\mathbf{q}} _{I} \( t \) - \bar{\mathbf{q}} _{J} \( t \) \)} + \sum _{K} \frac{\nabla _{\mathbf{q} _{j}} K _{h} \( \bar{\mathbf{q}} _{I} \( t \) - \bar{\mathbf{q}} _{K} \( t \) \)}{\sum _{J} K _{h} \( \bar{\mathbf{q}} _{J} \( t \) - \bar{\mathbf{q}} _{K} \( t \) \)} \]
  \nonumber
  \\
  &
  \qquad
  -
  \frac{1}{\zeta}
  \sum _{j = 1} ^{N - 1}
  A _{kj}
  \nabla _{\mathbf{q} _{j}} \Psi \( \bar{\mathbf{q}} _{I} \( t \) \)
  .
\end{align*}
We fix the number of representative particles $P = 1000$, and choose different time step sizes as $\delta _{t} \in \[ 1\e{-2}, 5\e{-3}, 2\e{-3}, 1\e{-3}, 5\e{-4} \]$.
We take $\delta _{t} ^{0} = 5\e{-4}$ as the benchmark and compare other cases with it.

To perform the comparison, we can calculate the difference of the particle position $\mathbf{q} _{t = 6, P = 1000} ^{\textup{ref}} \( \delta _{t} \)$ with respect to different time step size $\delta _{t}$. The relative time step error in $L ^{2}$ and $L ^{\infty}$ norms are,
\begin{align*}
  \textup{err} _{\textup{time, position}, L ^{2}} \( \delta _{t} \)
  & :=
  \frac{\lnm \mathbf{q} _{t = 6, P = 1000} ^{\textup{ref}} \( \delta _{t} \) - \mathbf{q} _{t = 6, P = 1000} ^{\textup{ref}} \( \delta _{t} ^{0} \) \rnm _{L ^{2}}}{\lnm \mathbf{q} _{t = 6, P = 1000} ^{\textup{ref}} \( \delta _{t} ^{0} \) \rnm _{L ^{2}}},
  \\
  \textup{err} _{\textup{time, position}, L ^{\infty}} \( \delta _{t} \)
  & :=
  \frac{\lnm \mathbf{q} _{t = 6, P = 1000} ^{\textup{ref}} \( \delta _{t} \) - \mathbf{q} _{t = 6, P = 1000} ^{\textup{ref}} \( \delta _{t} ^{0} \) \rnm _{L ^{\infty}}}{\lnm \mathbf{q} _{t = 6, P = 1000} ^{\textup{ref}} \( \delta _{t} ^{0} \) \rnm _{L ^{\infty}}}.
\end{align*}

Given the particle position $\mathbf{q} _{t = 6, P = 1000} ^{\textup{ref}} \( \delta _{t} \)$, we are able to recover the polymer configuration density function $f _{t = 6, P = 1000} ^{\textup{ref}} \( \mathbf{q}, \delta _{t} \)$ through the same kernel \autoref{eq:model_DeteParAppr} at any configuration $\mathbf{q} \in \mathbb{R} ^{\[ N - 1 \] \times d}$
\begin{align*}
  f _{t = 6, P = 1000} ^{\textup{ref}} \( \mathbf{q}, \delta _{t} \)
  :=
  \frac{1}{P} \sum _{I = 1} ^{P} K _{h} \( \mathbf{q} - \( \mathbf{q} _{t = 6, P = 1000} ^{\textup{ref}} \( \delta _{t} \) \) _{I} \),
\end{align*}
and the kernel bandwidth \autoref{eq:model_bandwidth} in the VDS.
In order to compare two polymer configuration density functions with different time step sizes, we generate a set of configurations $\mathbf{q} \in \lbk \mathbf{q} _{J} \rbk$, and calculate the difference of $f$ on this set with respect to $\ell ^{2}$ and $\ell ^{\infty}$ norms,
\begin{align*}
  \textup{err} _{\textup{time, density}, \ell ^{2}} \( \delta _{t} \)
  & :=
  \[ \frac{\sum _{J} \[ f _{t = 6, P = 1000} ^{\textup{ref}} \( \mathbf{q} _{J}, \delta _{t} \) - f _{t = 6, P = 1000} ^{\textup{ref}} \( \mathbf{q} _{J}, \delta _{t} ^{0} \) \] ^{2}}{\sum _{J} \[ f _{t = 6, P = 1000} ^{\textup{ref}} \( \mathbf{q} _{J}, \delta _{t} ^{0} \) \] ^{2}} \] ^{1 / 2},
  \\
  \textup{err} _{\textup{time, density}, \ell ^{\infty}} \( \delta _{t} \)
  & :=
  \frac{\max _{J} \lmdl f _{t = 6, P = 1000} ^{\textup{ref}} \( \mathbf{q} _{J}, \delta _{t} \) - f _{t = 6, P = 1000} ^{\textup{ref}} \( \mathbf{q} _{J}, \delta _{t} ^{0} \) \rmdl}{\max _{J} \lmdl f _{t = 6, P = 1000} ^{\textup{ref}} \( \mathbf{q} _{J}, \delta _{t} ^{0} \) \rmdl}.
\end{align*}
Specifically, we choose the set $\lbk \mathbf{q} _{J} \rbk$ on a mesh grid, where each component of $\mathbf{q} _{J}$ is taken from the set $\lbk -2, 0, 2 \rbk$. By checking the distribution, this set covers the major part of the support of $f$.

Another key quantity that can measure the difference in the results is the polymer stress in the discrete formula,
\begin{align*}
  \boldsymbol{\tau} _{t = 6, P = 1000} ^{\textup{ref}} \( \delta _{t} \) 
  := 
  \frac{n}{P} \sum _{I = 1} ^{P} \( \mathbf{q} _{t = 6, P = 1000} ^{\textup{ref}} \( \delta _{t} \) \) _{I} \nabla \Psi \( \( \mathbf{q} _{t = 6, P = 1000} ^{\textup{ref}} \( \delta _{t} \) \) _{I} \).
\end{align*}
The relative stress error in the Frobenius norm is,
\begin{align*}
  \textup{err} _{\textup{time, stress}, \textup{F}} \( \delta _{t} \)
  & :=
  \frac{\lnm \boldsymbol{\tau} _{t = 6, P = 1000} ^{\textup{ref}} \( \delta _{t} \) - \boldsymbol{\tau} _{t = 6, P = 1000} ^{\textup{ref}} \( \delta _{t} ^{0} \) \rnm _{\textup{F}}}{\lnm \boldsymbol{\tau} _{t = 6, P = 1000} ^{\textup{ref}} \( \delta _{t} ^{0} \) \rnm _{\textup{F}}}.
\end{align*}

\begin{figure}[!ht]
  \centering
  \includegraphics[width = 0.49\textwidth]{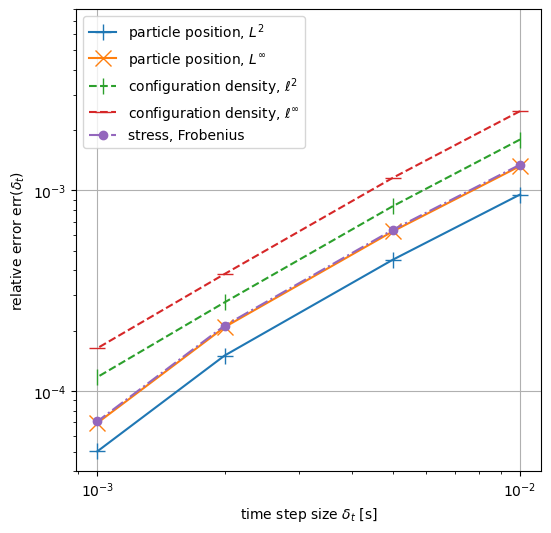}
  \includegraphics[width = 0.49\textwidth]{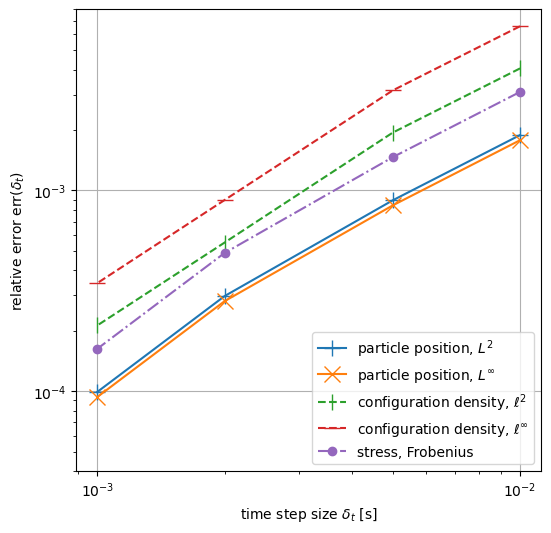}
  \caption{Various numerical error $\textup{err} _{\textup{time}} \( \delta _{t} \)$ vs. time step size $\delta _{t}$ in the reference dynamics. The left part shows the results from explicit schemes. Parameters: $4$-bead (homogeneous bonds) polymer molecules with simple shear flow, particle number $P = 1000$, reference time step size $\delta _{t} ^{0} = 5\e{-4}$. The right part shows the results from implicit schemes. Parameters: the same as the left part.}
  \label{fig:accuracy_explicit_dt}
\end{figure}

We numerically calculate the above relative errors in the left part of \autoref{fig:accuracy_explicit_dt}. It shows that the errors decay linearly as the time step size $\delta _{t}$ decreases. When $\delta _{t} = 1\e{-3}$, the numerical error is about $1\e{-4}$.
Compared with the POD-MOR error shown in \autoref{sec:experiments}, the error due to different time step sizes is much smaller.

Similarly, the first-order implicit scheme is applied to the reference model \autoref{eq:model_DymRepPar}
\begin{align*}
  &
  \quad
  \frac{1}{\delta _{t}} \[ \bar{\mathbf{q}} _{k, I} \( t ^{+} \) - \bar{\mathbf{q}} _{k, I} \( t \) \]
  +
  \mathbf{u} \cdot \nabla _{\mathbf{x}} \bar{\mathbf{q}} _{k, I} \( t \)
  -
  \bar{\mathbf{q}} _{k, I} \( t \) \cdot \nabla _{\mathbf{x}} \mathbf{u}
  \nonumber
  \\
  &
  =
  -
  \frac{\kB T}{\zeta}
  \sum _{j = 1} ^{N - 1}
  A _{kj}
  \[ \frac{\sum _{K} \nabla _{\mathbf{q} _{j}} K _{h} \( \bar{\mathbf{q}} _{I} \( t ^{+} \) - \bar{\mathbf{q}} _{K} \( t ^{+} \) \)}{\sum _{J} K _{h} \( \bar{\mathbf{q}} _{I} \( t ^{+} \) - \bar{\mathbf{q}} _{J} \( t ^{+} \) \)} + \sum _{K} \frac{\nabla _{\mathbf{q} _{j}} K _{h} \( \bar{\mathbf{q}} _{I} \( t ^{+} \) - \bar{\mathbf{q}} _{K} \( t ^{+} \) \)}{\sum _{J} K _{h} \( \bar{\mathbf{q}} _{J} \( t ^{+} \) - \bar{\mathbf{q}} _{K} \( t ^{+} \) \)} \]
  \nonumber
  \\
  &
  \qquad
  -
  \frac{1}{\zeta}
  \sum _{j = 1} ^{N - 1}
  A _{kj}
  \nabla _{\mathbf{q} _{j}} \Psi \( \bar{\mathbf{q}} _{I} \( t ^{+} \) \)
  ,
\end{align*}
which is a similar setup as \cite{bao2024micro, bao2025deterministic}.
For each time step, we solve the nonlinear equation by iteration.

Again, we numerically calculate the relative errors in the right part of \autoref{fig:accuracy_explicit_dt}. It also shows that the errors decay linearly as the time step size $\delta _{t}$ decreases. When $\delta _{t} = 1\e{-3}$, the numerical error is of the order of $1\e{-4}$.


Then, we compare the explicit and the implicit using their benchmark cases $\delta _{t} = 5\e{-4}$.
The particle position error in relative $L ^{2}$ and $L ^{\infty}$ norms are $9.51\e{-5}$ and $1.01\e{-4}$, the polymer configuration density error in relative $L ^{2}$ and $L ^{\infty}$ norms are $1.17\e{-4}$ and $1.80\e{-4}$, and the polymer stress error in relative Frobenius norm is $9.56\e{-5}$. That means the difference between these two benchmark cases is smaller than the errors shown in \autoref{fig:accuracy_explicit_dt}, so the choice of the explicit or implicit schemes is not that important at this point. 

Since solving the implicit method costs much more computation, the above results suggest we use an explicit scheme for the sake of efficiency. 
So in \autoref{sec:experiments}, we just focus on the explicit scheme for the POD-MOR, where the numerical error due to time integration is proved to be small enough.

We remark that the above convergence results in \autoref{fig:accuracy_explicit_dt} are based on the fixed number of representative particles $P = 1000$, but the convergence of the relative errors (particle position, polymer configuration density function, polymer stress) show similar plots when $P$ is in the range from $100$ to $10000$. Specifically, when increasing $P$, the relative errors at $\delta _{t} = 1\e{-3}$ show a slow increasing trend, but the magnitudes keep the value around $1\e{-4}$. That means the numerical error due to the time step size is not sensitive to the number of representative particles in the current setup.

Finally, we vary the number of particles. We fix the time step size $\delta _{t} = 1\e{-3}$ and choose different number of representative particles as $P \in \[ 100, 200, 500, 1000, 5000 \]$. We take $P ^{0} = 5000$ as the benchmark and compare the other cases with it.

Since we are not able to check the error of particle position due to the different number of particles, we consider the error of polymer configuration density function $f$ and polymer stress $\boldsymbol{\tau}$,
\begin{align*}
  \textup{err} _{\textup{poly-num, density}, \ell ^{2}} \( P \)
  & :=
  \[ \frac{\sum _{J} \[ f _{t = 6, \delta _{t} = 1\e{-3}} ^{\textup{ref}} \( \mathbf{q} _{J}, P \) - f _{t = 6, \delta _{t} = 1\e{-3}} ^{\textup{ref}} \( \mathbf{q} _{J}, P ^{0} \) \] ^{2}}{\sum _{J} \[ f _{t = 6, \delta _{t} = 1\e{-3}} ^{\textup{ref}} \( \mathbf{q} _{J}, P ^{0} \) \] ^{2}} \] ^{1 / 2},
  \\
  \textup{err} _{\textup{poly-num, density}, \ell ^{\infty}} \( P \)
  & :=
  \frac{\max _{J} \lmdl f _{t = 6, \delta _{t} = 1\e{-3}} ^{\textup{ref}} \( \mathbf{q} _{J}, P \) - f _{t = 6, \delta _{t} = 1\e{-3}} ^{\textup{ref}} \( \mathbf{q} _{J}, P ^{0} \) \rmdl}{\max _{J} \lmdl f _{t = 6, \delta _{t} = 1\e{-3}} ^{\textup{ref}} \( \mathbf{q} _{J}, P ^{0} \) \rmdl},
  \\
  \textup{err} _{\textup{poly-num, stress}, \textup{F}} \( P \)
  & :=
  \frac{\lnm \boldsymbol{\tau} _{t = 6, \delta _{t} = 1\e{-3}} ^{\textup{ref}} \( P \) - \boldsymbol{\tau} _{t = 6, \delta _{t} = 1\e{-3}} ^{\textup{ref}} \( P ^{0} \) \rnm _{\textup{F}}}{\lnm \boldsymbol{\tau} _{t = 6, \delta _{t} = 1\e{-3}} ^{\textup{ref}} \( P ^{0} \) \rnm _{\textup{F}}},
\end{align*}
where the terms $f _{t = 6, \delta _{t} = 1\e{-3}} ^{\textup{ref}} \( \mathbf{q} _{J}, P \)$ and $\boldsymbol{\tau} _{t = 6, \delta _{t} = 1\e{-3}} ^{\textup{ref}} \( P \)$ can be defined in the same way as before.

\begin{figure}[!ht]
  \centering
  \includegraphics[width = 0.5\textwidth]{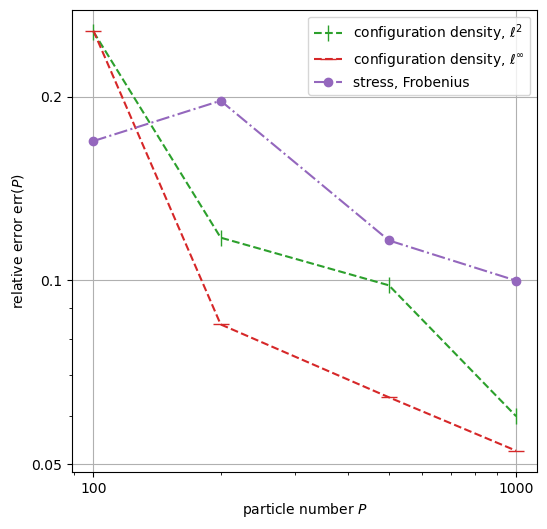}
  \caption{Various numerical error $\textup{err} _{\textup{poly-num}} \( P \)$ vs. number of representative particles $P$ in the reference dynamics. Parameters: $4$-bead (homogeneous bonds) polymer molecules with simple shear flow, reference particle number $P ^{0} = 5000$, time step size $\delta _{t} = 0.001$, explicit scheme.}
  \label{fig:accuracy_explicit_P}
\end{figure}

We numerically calculate the above relative errors in \autoref{fig:accuracy_explicit_P}. It shows that the errors decay as the number of particles $P$ increases. 

When $P = 1000$, the relative numerical error is about $5\% \sim 10\%$ (in different metrics), which is larger than the temporal error, so this value can be regarded as the numerical error of the reference dynamics, which is the benchmark of the POD-MOR error.

We should note that the decreasing rate of the relative polymer configuration density error in $L ^{2}$ norm seems slowing down when $P$ is approaching $1000$, which may be due to many issues, such as the set $\lbk \mathbf{q} _{J} \rbk$ we used for evaluating $f$. Such an issue is also shown in \cite{bao2024micro}.

By varying the parameters over a range of values, the above convergence test shows that our numerical scheme is convergent and stable with different time step sizes and polymer numbers, which confirms the robustness against discretization choices.

\end{document}